\documentclass[a4paper,final]{elsarticle}
\usepackage[a4paper,left=1in,right=1in,top=1in,bottom=1in,footskip=.25in]{geometry}

\usepackage{multirow}
\usepackage{float}
\usepackage{lipsum}
\usepackage{subcaption}
\usepackage{amsmath}
\usepackage{amsfonts}
\usepackage{upgreek}
\usepackage{textcomp}
\usepackage{lineno}
\usepackage{bm}
\usepackage{enumitem}
\usepackage[export]{adjustbox}

\usepackage{float}
\usepackage{algorithm}
\usepackage{algpseudocode}
\usepackage{algcompatible}
\biboptions{sort&compress}

\begin{document}

\begin{frontmatter}


\title{Fast Stochastic Subspace Identification of Densely Instrumented Bridges Using Randomized SVD}  

\journal{Mechanical Systems and Signal Processing}



\author[inst1]{Elisa Tomassini}
\ead{elisa.tomassini@dottorandi.unipg.it}

\author[inst2]{Enrique Garc\'{i}a-Mac\'{i}as}
\ead{enriquegm@ugr.es}
\cormark[1]

\author[inst1]{Filippo Ubertini}
\ead{filippo.ubertini@unipg.it}

\address[inst1]{Department of Civil and Environmental Engineering, University of Perugia. Via G. Duranti, 93 - 06125 Perugia, Italy.}

\address[inst2]{Department of Structural Mechanics and Hydraulic Engineering, University of Granada, Av. Fuentenueva sn, 18002 Granada, Spain.}



\begin{abstract}
The rising number of bridge collapses worldwide has compelled governments to introduce predictive maintenance strategies to extend structural lifespan. In this context, vibration-based Structural Health Monitoring (SHM) techniques utilizing Operational Modal Analysis (OMA) are favored for their non-destructive and global assessment capabilities. However, long multi-span bridges instrumented with dense arrays of accelerometers present a particular challenge, as the computational demands of classical OMA techniques in such cases are incompatible with long-term SHM. To address this issue, this paper introduces Randomized Singular Value Decomposition (RSVD) as an efficient alternative to traditional SVD within Covariance-driven Stochastic Subspace Identification (CoV-SSI). The efficacy of RSVD is also leveraged to enhance modal identification results and reduce the need for expert intervention by means of 3D stabilization diagrams, which facilitate the investigation of the modal estimates over different model orders and time lags. The approach's effectiveness is demonstrated on the San Faustino Bridge in Italy, equipped with over 60 multiaxial accelerometers.
\end{abstract}




\end{frontmatter}

\section{Introduction}

Improving bridge safety has become a key priority on the political agenda due to the increasing number of bridge failures worldwide in recent years. This trend is largely attributed to the fact that much of the infrastructure in Europe, built in the mid-20th century, is now approaching or exceeding its intended service life of 50-100 years. Indeed, approximately 33\% of European highway bridges are estimated to have structural deficiencies~\cite{Europe2019,cuerva2021actual}. For instance, the German Federal Highway Research Institute (BASt)~\cite{fur2019zustandsnoten} reported in 2019 that 12.2\% of nearly 40,000 national road bridges were significantly deteriorated. In March 2024, the French Senate~\cite{FranceRT} reported that 25\% (9,000 structures) of national bridges have severe structural deficiencies and require immediate interventions, allocating €35 million for repairs. In this context, it is critical to implement effective preventive maintenance strategies to prevent bridge collapses, tailor maintenance interventions to maximize their lifespan, and prioritize resources to manage such a massive infrastructure stock. Notably, inadequate or untimely maintenance is recognized as one of the key reasons for bridge collapses as reported by Zhang \textit{et al.}~\cite{zhang2022}. This has prompted governments worldwide to boost investments and implement new regulations to transition from traditional and inefficient maintenance strategies--relying exclusively on periodic visual inspections--to more effective Structural Health Monitoring (SHM) approaches. In this regard, Italy represents an exceptional example which, after the dramatic collapse of the Morandi bridge causing 45 fatalities in 2018~\cite{calvi2019once}, enacted the Guidelines on Risk Classification and Management~\cite{GUIDA2020}. Today, these new regulations are among the most advanced standards in the world, recognizing the crucial role of SHM in addressing the challenge of ageing infrastructure.  

Among the wide variety of available SHM approaches, vibration-based SHM techniques through Operational Modal Analysis (OMA) are particularly popular owing to their non-destructive and non-intrusive nature, global damage identification capabilities, and compatibility with permanent monitoring~\cite{hasani2023operational}. These techniques exploit ambient accelerations recorded at strategic locations on a structure to estimate its modal characteristics (i.e.~resonant frequencies, mode shapes, and damping ratios), which can be used as damage-sensitive features given their dependence on the stiffness and energy dissipation properties of the structure \cite{GarciaMacias2020a,Quqa2021,Yun2021}. Therefore, the appearance of persistent variations in these modal features may signal the onset of structural damage. In this light, considerable research efforts have been devoted over the past decades to the development of diverse time-, frequency-, and time-frequency-domain OMA techniques (refer e.g. to~\cite{brincker2015introduction,Guan2019,zahid2020review} for an extensive bibliographic review). Most recent efforts have focused on developing automated OMA procedures for continuous SHM~\cite{Rainieri2010,he2024review}, particularly those based on the automated interpretation of stabilization diagrams from time-domain Stochastic Subspace Identification (SSI)~\cite{peeters1999}, either in the covariance-driven (CoV-SSI)~\cite{garcia2020mova} or data-driven (Data-SSI)~\cite{Zini2022} versions. These techniques commonly operate by fitting various state-space models across different model orders and plotting them in a model order versus eigenfrequency diagram, known as a stabilization diagram. Based on this diagram, physical modes can be discriminated from mathematical artefacts caused by model-order over-specification, which appear as vertical lines of stable poles. To automate this process, the identified poles are typically first filtered by a set of Hard Criteria (HC) and Soft Criteria (SC), which help distinguish between potentially physical poles and clearly spurious ones~\cite{reynders2012system}. Some of the most effective indicators for HC/SC include Modal Phase Collinearity (MPC)~\cite{pappa1993consistent}, Modal Transfer Norm (MTN)~\cite{reynders2008reference}, Modal Coherence Indicator~\cite{lardies2011modal}, Mean Phase Deviation (MPD)~\cite{dederichs2023experimental}, and uncertainty bounds on modal parameters~\cite{reynders2008uncertainty}. After cleansing the stabilization diagram, identifying vertical lines of stable poles becomes easier. To this purpose, a number of partitioning or clustering methods have been proposed in the literature, including, among others, hierarchical clustering~\cite{garcia2022p3p}, histogram analysis~\cite{romanazzi2023iterative}, fuzzy C-means~\cite{Bezdek1981fuzzycmean,Babuska2000fuzzycmean,Charbonnel2021}, Gaussian Mixture Models (GMM)~\cite{Cheema2021}, density-based clustering~\cite{civera2023dbscan}, and blind k-means~\cite{Tronci2022}. Such techniques automate the modal identification of structures, enabling the tracing of time series of modal signatures through a modal tracking approach~\cite{garcia2022integrated,HE2022int}, and subsequently facilitating damage assessment. However, widespread adoption of these techniques for SHM of infrastructural systems at a regional scale is hindered by their inefficiency. On one hand, they require expert intervention to set identification parameters and, on the other hand, they demonstrate low computational efficiency when managing dense sensor networks.

The success of these techniques has motivated infrastructure managers to instrument many bridges within the transportation network. A notable example is the SHM program (\textit{Programma SHM}~\cite{ProgrammaSHM}) implemented in 2022 by Anas S.p.A., the largest public infrastructure manager in Italy. The program aims to deploy continuous vibration-based SHM systems in 1,000 bridges across Italy by 2026, with an overall budget of €275 million. These bridges are often densely instrumented with wired accelerometers connected to data acquisition systems, which store the recordings on remotely controlled onboard computers. Moreover, a significant portion of the bridge stock features isostatic or quasi-isostatic schemes such as long multi-span simply supported bridges or long continuous girder bridges with half-joints. The dynamic identification of these bridges using automated OMA poses a twofold challenge: (i) substantial expert intervention and parameter tuning, and (ii) high computational burden along with the need for high-performance onboard hardware. To alleviate these bottlenecks, analysts often choose to employ parallel processing of different subgroups of channels based on the structure's static scheme (i.e.~spanwise analyses). While this method effectively reduces computational demand, Tomassini \textit{et al.}~\cite{tomassini2023model} proved that it may compromise the ability to perform a comprehensive assessment of the entire structure, as such bridges typically exhibit weak coupling between adjacent spans.

To address the limitations of current automated OMA approaches, considerable efforts have been made in recent years to harness the potential of machine learning and Artificial Intelligence (AI) in developing next-generation SHM systems. For instance, it is worth noting the work by Liu and co-authors~\cite{liu2023machine}, who developed a two-level AI algorithm for SSI. The algorithm includes a model order determination neural network (NN), followed by a second modal identification NN. Their approach showcased comparable accuracy to classical SSI in a real-world suspension bridge, achieving remarkable reductions in computational burden. Similarly, Shim \textit{et al.}~\cite{shim2023ssi} presented an SSI-based long-short term memory (SSI-LSTM) method to track variations in the modal parameters of structures. The developed AI model takes raw acceleration data as input and is trained to reproduce the modal estimates obtained by CoV-SSI, enabling quasi real-time estimation of the evolution of the modal signatures. Another noteworthy contribution was made by Lin \textit{et al.}~\cite{liu2021machine}, who encapsulated the principles of blind source separation (BSS) into the loss function of a self-coding NN, yielding promising results in a real-world steel box-girder cable-stayed bridge. Shu and co-authors~\cite{shu2023multi} developed a multi-task DNN for the automated identification of independent modes extracted from SCA, achieving comparable accuracy levels to FDD and CoV-SSI with critically lower computational time demands. In their approach, mode shapes are directly extracted as the weights between the last two layers of neurons in the NN, while the resonant frequencies and damping ratios are estimated from the independent components using the random decrement technique (RDT). Similarly, Hern{\'a}ndez-Gonz{\'a}lez \textit{et al.}~\cite{hernandez2024ai} recently presented a Multitask Learning Deep Neural Network (MTL-DNN) for fast and automated BSS of structures, facilitating the extraction of complex-valued mode shapes with minimal computational demands. Despite these successful experiences, a major drawback arises from the various user-defined hyperparameters in AI and the complexities involved in their training. This includes the selection of the modal parameters in the OMA method used for training the AI algorithms, which hampers their scalability to large-scale infrastructural networks. 

Given the aforementioned difficulties, there remains significant interest in optimizing classical SSI techniques. In this vein, Ubertini and co-authors~\cite{Ubertini2013} proposed using hierarchical clustering over the poles extracted by Data-SSI considering a range of block rows for constructing the Hankel matrix, thus avoiding the need to fix this parameter. Rainieri and Fabbrocino~\cite{rainieri2014influence} proposed the use of convergence analysis of the modal estimates to identify the optimal number of block rows of the Hankel/Topelitz matrices. Similarly, a recent work by Rosso \textit{et al.}~\cite{rosso2023intelligent} developed a random forest algorithm to predict the optimal combination of the control parameters in CoV-SSI, including the number of block rows in the Toeplitz covariance matrix and the maximum model order. While these approaches alleviate the need for expert intervention in selecting the control parameters, the computational cost in the identification remains a limitation. Specifically, the most computationally intensive step in CoV-SSI and Data-SSI is the Singular Value Decomposition (SVD) of the block Toeplitz matrix of output covariance matrices or the Hankel matrix of the measurements, respectively. This is particularly critical for densely instrumented structures, as the size of the matrices increases linearly with the number of measurement channels. In this light, some authors have attempted to alleviate the computational burden in this decomposition. A noticeable contribution was made by Peeters and De Roeck~\cite{peeters1999reference}, who developed the so-called reference-based SSI algorithm (SSI/ref). Inspired by multi-setup OMA, this approach takes a set of channels as references, projecting the row space of the future outputs in the Hankel matrix of the Data-SSI algorithm into the row space of the past outputs of the reference sensors. Later extended to the case of combined deterministic-stochastic SSI by Reynders and De Roeck~\cite{reynders2008reference}, SSI/ref has demonstrated a considerable reduction in computation time by reducing the dimension of the matrices involved in the SVD. Nonetheless, the reference sensors must be properly chosen to avoid quality loss in the estimated modal parameters, and the accuracy of the displacements for the non-reference channels is often lower. Alternatively, Sun \textsl{et al.}~\cite{sun2023fast} recently proposed using a fast truncated SVD decomposition to reduce the computational cost of CoV-SSI. Their method involves transforming the SVD step into a truncated eigendecomposition of a symmetric matrix, followed by a further transformation into the eigendecomposition of a tridiagonal matrix using the finite-precision Lanczos algorithm. Through a 9-degree-of-freedom (DOF) benchmark example, their approach demonstrated reductions in computational time of up to approximately 97\% with almost negligible errors in the SVD.

Given the lack of generalized SSI algorithms that offer both low computational cost and minimal expert intervention in defining the control parameters, this work introduces a novel CoV-SSI algorithm based on randomized SVD (RSVD). In recent decades, Randomized numerical linear algebra has emerged as a powerful technique for reducing large matrices to lower-order representations without losing crucial information~\cite{woolfe2008,drineas2016,Haiko2011}. It has found applications in various computationally intensive fields, including machine learning~\cite{rokhlin2018}, fluid dynamics~\cite{martini2021}, medical ultrasound imaging~\cite{song2017accelerated}, and computational mechanics~\cite{cao2013svd}, among others. Leveraging this potential, Wei \textsl{et al.}~\cite{Wei2024} recently proposed the implementation of RSVD within the CoV-SSI algorithm, referred to as the lightweight Stochastic Subspace Identification (lwSSI) method. Specifically, lwSSI combines RSVD with sliding data windows to track time-varying modal characteristics and iteratively eliminate spurious modes, thereby enhancing both computational speed and modal estimation precision. The effectiveness of lwSSI was demonstrated on a numerical finite element model of a gravity dam simulating ambient vibrations recorded with 14 sensors, as well as on a laboratory cantilever beam instrumented with 7 accelerometers and excited by a shaker. Despite lwSSI's proven computational advantages, the full potential of RSVD remains partially unexplored. In particular, RSVD's capabilities could be especially impactful when applied to large, densely instrumented structures, where computational demands are high and parameter selection for Cov-SSI is challenging, often requiring significant expert intervention. To address these limitations, this paper further investigates the use of RSVD, not only to streamline computations, but also to enhance identification accuracy and reduce the need for expert intervention. Notably, RSVD is used as a substitute for SVD in the CoV-SSI algorithm, and parametric analyses are conducted to identify the optimal control parameters that maintain results accuracy while minimizing computational burden. On the other hand, although traditional CoV-SSI involves iterating across different model orders and presenting stable poles in a 2D stabilization diagram, the final results are also highly sensitive to the time lag, which is typically kept constant based on rule-of-thumb criteria or the analyst's previous experience. To eliminate the uncertainty associated with this control parameter and minimize expert intervention, RSVD's ability to significantly reduce computational burden is leveraged to enhance the modal identification process. This enables the inexpensive creation of 3D stabilization diagrams and the investigation of modal pole stability not only across various model orders but also across different time lag values. Finally, using a 2-stage clustering approach, the physical modal signatures are automatically extracted from the 3D stabilization diagrams. The effectiveness of the proposed approach is evaluated through two case studies: a numerical 10-DOFs dynamic system and a real-world in-operation bridge. The second case study involves the San Faustino Bridge in Perugia, Italy -- a nine-span continuous box girder bridge with 10 half-joints interrupting the continuity of the deck. This bridge is instrumented with over 50 accelerometers, totalling 114 channels arranged on both the deck and the piers, representing a genuine example of a densely instrumented bridge.

The remainder of the paper is organized as follows: Section \ref{section2} outlines the theoretical background of modal identification through CoV-SSI, RSVD, and the methodology for generating the 3D stabilization diagram. Section \ref{section3} presents the numerical results and discussion. Finally, Section \ref{section4} concludes the paper.


\section{Theoretical background and methodology}\label{section2}

This section is structured to provide a thorough foundation for understanding the methodologies presented in this study. Initially, Section~\ref{section21} outlines the key theoretical principles of the CoV-SSI algorithm. Next, Section~\ref{section22} explores the theoretical framework of RSVD and its potential as an alternative to traditional SVD. Finally, Section~\ref{section23} offers a detailed explanation of the procedural steps necessary for generating the proposed 3D stabilization diagram.

\subsection{Covariance-drive Stochastic Subspace Identification (CoV-SSI)}\label{section21}

The differential equation of equilibrium of a dynamic system characterized by $n$ DOFs is completely defined by the continuous state-space equations as follows:
\begin{equation}  \label{Eq.SSM}
\begin{aligned}
&\textrm{State equation:} &\dot{\textbf{x}} (t) &= \textbf{A}_\textrm{C} \, \textbf{x} (t) + \textbf{B}_\textrm{C}\, \textbf{u} (t),\\
&\textrm{Observation equation:} &\dot{\textbf{y}} (t) &= \textbf{C}_\textrm{C}\, \textbf{x} (t) + \textbf{D}_\textrm{C}\, \textbf{u} (t).
\end{aligned}
\end{equation}
\noindent The \textit{state equation} in Eq.~(\ref{Eq.SSM}) defines the relationship between the displacement and velocities of the dynamic system contained in the state vector $\textbf{x}(t) \in \mathbb{R}^{2n}$, and the $n_i$ applied loads $u_i(t) \in \mathbb{R}^{n_i}$. The state matrix  $\textbf{A}_\textrm{C}\in \mathbb{R}^{2n \times 2n}$ contains mechanical information of the dynamic system, and the input matrix $ \textbf{B}_\textrm{C}\in \mathbb{R}^{2n \times 2n}$ relates the applied loads to the respective DOFs. The first term in the second equation in Eq.~(\ref{Eq.SSM}) is the observation vector $\textbf{y}(t) \in \mathbb{R}^{l}$, which contains a subset of $l$ measured outputs of the system. Therefore, the \textit{observation equation} establishes the relationship between the available output measurements in the structure $\textbf{y}(t)$ and the state vector $\textbf{x}(t)$, through the observation matrix $ \textbf{C}_\textrm{C}\in \mathbb{R}^{l \times 2n}$ and the direct transmission matrix  $ \textbf{D}_\textrm{C}\in \mathbb{R}^{l \times n_i}$.

The modal characteristics of the dynamic system can be derived from the eigenvalue decomposition of the state matrix $\textbf{A}_\textrm{C}$ as \cite{rainieri2014operational}:
\begin{equation}  \label{Eq.freqmod}
\begin{aligned}
&\textbf{A}_\textrm{C}= \bm{\Psi} \bm{\Lambda}_\textrm{C}\bm{\Psi}^{-1}, & \\
&\text{where:}  \quad \bm{\Lambda}_\textrm{C}=
\begin{bmatrix}
\bm{\Lambda} & 0 \\
0 & \bm{\Lambda}^*
\end{bmatrix},  \quad \bm{\Psi} =
\begin{bmatrix}
\bm{\Theta} & \bm{\Theta}^* \\
\bm{\Theta} \bm{\Lambda} & \bm{\Theta}^* \bm{\Lambda}^*
\end{bmatrix}, \quad \bm{\Lambda} =
\begin{bmatrix}
\lambda_1 & & \\
& \ddots & \\
& & \lambda_{n}
\end{bmatrix}, \quad \bm{\Theta} =
\begin{bmatrix}
\bm{\phi}_1 & \cdots & \bm{\phi}_{n}
\end{bmatrix}, &
\end{aligned}
\end{equation}
\noindent where $\bm{\Lambda}_C$ and $\bm{\Psi}$ represent the eigenvalue and eigenvector matrices, respectively, an the subscript * denotes the complex conjugate operator. The resonant angular frequencies $\omega_i$ and the modal damping ratios $\xi_i$, $i=1,\dots,n$ are related to the complex eigenvalues $\lambda_i$ through the following equation:
\begin{equation}
    \lambda_i = -\xi_i \omega_i + \textrm{i} \ \omega_i \sqrt{1 - \xi_i^2},
\end{equation}
\noindent while the observable modal matrix $\bm{\Phi} \in \mathbb{R}^{l \times n}$ derive from Eq.~(\ref{Eq.freqmod}) as:
\begin{equation}
    \bm{\Phi} = \textbf{C}_\textrm{C}\bm{\Psi}.
\end{equation}

Nonetheless, analog signals are digitized in practical applications through analogue-to-digital converters, resulting in discrete-time information. This process involves sampling the signals at a specific frequency  $f_s = 1/\Delta t$ , where  $\Delta t$  denotes the sampling interval. Consequently, the continuous-time functions  $\textbf{x}(t)$  and  $\textbf{y}(t)$ in Eq.~(\ref{Eq.SSM}) are replaced by their discrete-time counterparts $ \textbf{y}_h = \textbf{y}(h \Delta t)$ and $ \textbf{x}_h = \textbf{x}(h \Delta t)$ at time instants $h \Delta t$, where  $h$  is an integer. Moreover, within the framework of OMA, the system inputs $\textbf{u}(t)$  are unknown and so the ambient vibrations are modeled as white noise stochastic processes. Consequently, the discrete-time stochastic state-space model is formulated as follows:
\begin{equation}  \label{Eq.DSSM}
\begin{aligned}
\textbf{x}_{h+1} &= \textbf{A} \textbf{x}_h + \textbf{w}_h, \\
\textbf{y}_h &= \textbf{C} \textbf{x}_h + \textbf{v}_h,
\end{aligned}
\end{equation}
\noindent where $ w_h \in \mathbb{R}^{2n} $ and $ v_h \in \mathbb{R}^l $ represent white noise processes accounting for the effect of unknown inputs modeled as stochastic processes, model inaccuracies, and measurement noise.

On this basis, the CoV-SSI algorithm method identifies the stochastic discrete-time state-space model in Eq.~(\ref{Eq.DSSM}) from the output correlation matrix. Considering positive time intervals ranging from \( \Delta t \) to \( (2j_b - 1) \Delta t \), where $j_b$ is the time-lag step, the output correlation matrices \( \textbf{R}_1 \in \mathbb{R}^{l \times l} \) to \( \textbf{R}_{2j_b-1} \in \mathbb{R}^{l \times l} \) can be directly derived from the acceleration records.The resulting output correlation matrices can be arranged in a block-Toeplitz matrix form as \cite{magalhaes2011explaining}:
\begin{equation} \label{Eq.Toeplitz}
T_{1|j_b} = 
\begin{bmatrix}
\textbf{R}_{j_b} & \textbf{R}_{j_b-1} & \cdots & \textbf{R}_1 \\
\textbf{R}_{j_b+1} & \textbf{R}_{j_b} & \cdots & \textbf{R}_2 \\
\vdots & \vdots & \ddots & \vdots \\
\textbf{R}_{2j_b-1} & \textbf{R}_{2j_b-2} & \cdots & \textbf{R}_{j_b}
\end{bmatrix}
\in \mathbb{R}^{j_b l \times j_b l}.
\end{equation}
Given the factorization property of the correlation matrix \( \textbf{R}_j = \textbf{C} \textbf{A}^{j-1} \textbf{G} \) \cite{rainieri2014operational}, with  \textbf{G} being the next state-output correlation matrix (i.e. $ \textbf{G} = \textbf{C} \textbf{A}^{j-1} \mathbb{E}[x_{h+1} y_h^T]$), the Toeplitz matrix in Eq.~(\ref{Eq.Toeplitz}) can be decomposed as:
\begin{equation}\label{Eq.Toeplitz2}
 \textbf{T}_{1|j_b} = \begin{bmatrix} \textbf{C} \\ \textbf{CA} \\ \vdots \\ \textbf{CA}^{j_b-1} \end{bmatrix} 
 \begin{bmatrix} \textbf{A}^{j_b-1} \textbf{G} & \cdots & \textbf{AG} & \textbf{G} \end{bmatrix}
 = \textbf{O} \bm{\Gamma},
\end{equation}
\noindent with $\textbf{O}$ and $\bm{\Gamma}$ denoting the observability and controllability matrices, respectively. Furthermore, the singular value decomposition of the Toeplitz matrix reads:
\begin{equation}\label{Eq.SVD}
\textbf{T}_{1|j_b} = \textbf{U} \textbf{S} \textbf{V}^\textrm{T} = \begin{bmatrix}
\textbf{U}_1 & \textbf{U}_2\\
\end{bmatrix} \begin{bmatrix}
\textbf{S}_1 & 0 \\
0 & 0
\end{bmatrix} \begin{bmatrix}
\textbf{V}_1 \\
\textbf{V}_2 
\end{bmatrix} = \textbf{U}_1 \textbf{S}_1 \textbf{V}_1^\textrm{T},
\end{equation}
\noindent where $\textbf{U}_1 \in \mathbb{R}^{j_bl \times n_o}$ and $\textbf{V}_1 \in \mathbb{R}^{j_bl \times n_o}$ are the left and right singular vector matrices associated to the singular values matrix of the system $\textbf{S}_1 \in \mathbb{R}^{n_o \times n_o}$, where $n_o$ represents the number of non-zero singular values on the diagonal of $\textbf{S}$. Therefore,  $n_o$ also represents the size of the state matrix $\textbf{A}$. Comparing Eq.~(\ref{Eq.Toeplitz2}) and (\ref{Eq.SVD}), one can readily partition the observability $\textbf{O}$ and controllability matrices $\bm{\Gamma}$ as :
\begin{equation}
\begin{aligned}
\textbf{O} &= \textbf{US}^{1/2}, \\
\bm{\Gamma} &= \textbf{S}^{1/2}  \textbf{V}^\textrm{T}.
\end{aligned}
\end{equation}
\noindent Once the observability and controllability matrices are determined, the system matrices $\textbf{A}$ and $\textbf{C}$ can be extracted by mean of Eq.~(\ref{Eq.Toeplitz2}). Indeed, $\textbf{C}$ can be extracted from the first $l$ rows of $\textbf{O}$, while $\textbf{A}$ can be estimated as the least squares problem expressed by the following equation \cite{magalhaes2011explaining}:
\begin{equation}
\begin{bmatrix}
\textbf{C}  \\
\textbf{CA} \\
\vdots\\
\textbf{CA}_{j_b-1}
\end{bmatrix}
\textbf{A} =
\begin{bmatrix}
\textbf{CA}  \\
\textbf{CA}^2 \\
\vdots\\
\textbf{CA}_{j_b-1}
\end{bmatrix} \,
\Leftrightarrow \,
\textbf{A} =
\begin{bmatrix}
\textbf{C} \\
\textbf{CA}  \\
\vdots  \\
\textbf{CA}_{j_b-2} 
\end{bmatrix}^{\dag}
\begin{bmatrix}
\textbf{CA}  \\
\textbf{CA}^2 \\
\vdots\\
\textbf{CA}_{j_b-1}
\end{bmatrix}
=
{\textbf{O}{^{to^{\dag}}}}  \textbf{O}^{bo},
\end{equation}
\noindent where $\textbf{O}^{\text{to}}$ and $\textbf{O}^{\text{bo}}$ contain the first and the last $n_o\left(j_b-1\right)$ rows of $\textbf{O}$, respectively, while the symbol $^{\dag}$ represents the Moore–Penrose pseudo-inverse operator.

\noindent It can be demonstrated that the eigenvectors of matrix $\textbf{A}$ coincide with those of the continuous counterpart $\textbf{A}_C$ and that $\textbf{C} \equiv \textbf{C}_C$. Therefore, for a specific system order $n_2$, $n_2 \leq n_o$ (i.e., the number of retained singular values in Eq. (\ref{Eq.SVD})), the modal matrix $\bm{\Psi} \in \mathbb{C}^{l \times n_2}$ and the eigenvalues matrix $\textbf{M} \in \mathbb{R}^{n_2 \times n_2}$ of the discrete-time state-space model can be evaluated from the eigenvalue decomposition of $\textbf{A}$ as:
\begin{equation}
\textbf{A} = \Psi \textbf{M} \Psi^\textrm{T}, \, \textrm{with} \; \Psi = \begin{bmatrix} \bm{\psi}_1 & \cdots & \bm{\psi}_{n_2} \end{bmatrix} ,   \textbf{M} = \begin{bmatrix} \mu_1 & & \\ & \ddots & \\ & & \mu_{n_2} \end{bmatrix}.
\end{equation}
Finally, the corresponding eigenvalues $\lambda_h$, natural frequencies $f_h$, damping ratios $\xi_h$, and modal matrix $\Phi \in \mathbb{C}^{l \times n_2}$ of the continuous system can be evaluated as \cite{magalhaes2011explaining}:
\begin{equation}
\lambda_h = \frac{\ln \mu_h}{\Delta t}, \; f_h = \frac{\operatorname{Re} (\lambda_h)}{2 \pi \xi_h}, \; \xi_h = \frac{\operatorname{Re}(\mu_h)}{|\mu_h|}, \; \Phi = \textbf{C} \bm{\Psi}.
\end{equation}

The quality of the modal identification results is significantly influenced by the selection of the control parameters of the algorithm. The main parameters include the minimum and maximum model orders ($n_2 \in \mathbb{M}, \mathbb{M}=[N_{min}, N_{max}]\in \mathbb{N}$) and the time-lag $\tau=(2j_b-1)\Delta t$, which is often fixed following a rule-of-thumb or the previous experience by the analyst. Typically, the poles obtained by iteratively retaining $n_2 \in [N_{min}, N_{max}]$ in Eq. \ref{Eq.SVD} are plotted in a two-dimensional stabilization diagram on the frequency-model order plane. Afterwards, the poles are filtered to assess their stability along the model orders axis by applying both SC and HC as elaborated by Reynders \textit{et al.} in~\cite{reynders2011automated}.
The former involve the "hard" removal of poles that violate certain defined thresholds, such as maximum damping ratios ($\xi_{\textrm{max}}$), minimum Modal Phase Collinearity (MPC$_{\textrm{min}}$), and maximum Mode Phase Deviation (MPD$_{\textrm{max}}$). MPC~\cite{pappa1993consistent} assesses the linear relationship between the real and imaginary parts of the modal components, while MPD~\cite{dederichs2023experimental} evaluates the statistical variation of the phase angles of each modal component. Both parameters are dimensionless and range from 0 to 1.
Soft criteria (SC), on the other hand, are applied between every pair of poles $i$ and $j$ identified at consecutive model orders $m$ and $m-1$. These criteria consider poles as spurious if they violate thresholds related to relative differences in resonant frequencies $|f_i^m - f_j^{m-1}|/\textrm{max}(f_i^m, f_j^{m-1})$, damping ratios $|\xi_i^m - \xi_j^{m-1}|/\textrm{max}(\xi_i^m, \xi_j^{m-1})$, and Modal Assurance Criterion (MAC)~\cite{allemang2012MAC} values $\textrm{MAC}(\bm{\phi}^{m}_i, \bm{\phi}^{m-1}_j)$.
Poles classified as stable are then grouped using clustering algorithms to identify alignments in the stabilization diagram. To this aim, various algorithms have demonstrated effectiveness in the literature, as previously mentioned in the introduction. Among these, agglomerative hierarchical clustering techniques are widely used due to their insensitivity to initial conditions, ease of implementation, and straightforward interpretation of the hierarchical relationship between clusters. There are also multi-stage clustering methods, such as the one proposed in~\cite{Charbonnel2021}, which combines hierarchical clustering with fuzzy C-means clustering, effectively bypassing the need of applying SC to analyse the stability of the poles. Thus, a careful selection of all parameters involved in these steps is essential to achieve accurate identification results through the SSI-Cov algorithm.

\subsection{Randomized Singular Value Decomposition (RSVD)}\label{section22}

Randomized Numerical Linear Algebra (RandNLA) is a field within randomized mathematics that includes various algorithms designed to tackle fundamental problems like matrix multiplication, least squares regression, and eigenvalue decomposition using randomization techniques~\cite{Haiko2011, Mahoney2011}. These algorithms use random projections and subsampling to retain the key features of the original data, allowing for accurate approximations while significantly reducing computational complexity and memory usage.

Let us consider a matrix $\textbf{A} \in \mathbb{R}^{m \times n}$. A natural low-rank approximation of $\textbf{A}$ can be expressed in the form \cite{Haiko2011}:
\begin{equation}\label{Eq.apA}
    \textbf{A} \approx \textbf{QQ}^\textrm{T}\textbf{A},
\end{equation}
\noindent where $\textbf{Q} \in \mathbb{R}^{m \times k}$ represents an approximate basis for the range of $\textbf{A}$ with orthonormal columns and $k\leq n$ representing the numerical rank of $\textbf{A}$. Low-rank approximation techniques naturally lead to the discussion of RSVD, a matrix factorization method that employs random sampling to efficiently approximate SVD of a matrix. Unlike classical deterministic methods, RSVD utilizes random projections to reduce the dimensionality of the data prior to performing SVD, which significantly lowers the computational cost. This approach is particularly beneficial for large, sparse, or structured matrices where traditional SVD would be computationally prohibitive. The fundamental concept of RSVD is to construct a smaller, representative subspace of the original matrix through random projections, facilitating efficient approximation of its SVD. Therefore, the first step involves constructing a random matrix $\bm{\Omega} \in \mathbb{R}^{n \times k}$ with the purpose of projecting the high-dimensional data $\textbf{A}$ into a lower-dimensional subspace~\cite{Haiko2011}:

\begin{equation}\label{eq:randomsamplingA}
\textbf{Y} = \textbf{A} \bm{\Omega},
\end{equation}

\noindent where $\textbf{Y} \in \mathbb{R}^{m \times k}$ is the sampled matrix of $\textbf{A}$. Among the different random distributions that can be used to construct matrix $\bm{\Omega}$, this work adopts a matrix with normally distributed random entries, as recommended by several authors in the literature (see e.g.~\cite{Haiko2011, Martinsson2011}). Therefore, the matrix $\textbf{Q}$ can be derived from the QR decomposition~\cite{strang2019Linear_Alg_Learning_data} of $\textbf{Y}=\textbf{QR}$ leveraging the computation of a reduced matrix $\textbf{P}$:

\begin{equation}\label{Eq.B}
\textbf{P} = \textbf{Q}^\textrm{T} \textbf{A}.
\end{equation}

The matrix $\textbf{P} \in \mathbb{R}^{k \times n}$ retains the essential characteristics of $\textbf{A}$ but with significantly reduced dimensions. Performing SVD on $\textbf{P}$:

\begin{equation}
\textbf{P} = \tilde{\textbf{U}} \bm{\Sigma} \textbf{V}^\textrm{T}
\end{equation}

\noindent provides the singular values $\bm{\Sigma} \in \mathbb{R}^{k \times k}$ and singular vectors $\tilde{\textbf{U}} \in \mathbb{R}^{k \times k}$ and $\textbf{V} \in \mathbb{R}^{n \times k}$. The comparison between Eqs.~(\ref{Eq.apA}) and (\ref{Eq.B}) yields to the low-rank factorization:

\begin{equation}
\textbf{A} \approx \textbf{Q P}.
\end{equation}

Finally, the RSVD of the original matrix $\textbf{A}$ can be reconstructed as~\cite{Haiko2011}:

\begin{equation}\label{eq:reconstruct_A}
\textbf{A} \approx \textbf{U} \bm{\Sigma} \textbf{V}^\textrm{T},
\end{equation}

\noindent where $\textbf{U} = \textbf{Q} \tilde{\textbf{U}}, \textbf{U} \in \mathbb{R}^{m \times k}$.

The resulting algorithm is summarized in Algorithm \ref{alg:rsvd}. The power of RSVD lies in its ability to dramatically reduce computational complexity. Traditional SVD requires $O(mn \min(m, n))$ operations, which turns prohibitive for large matrices. In contrast, RSVD reduces this to $O(mnk)$, where $k$ is the target rank, typically much smaller than $m$ or $n$. This reduction is achieved without significant loss of accuracy, making RSVD particularly well-suited for applications involving large datasets.

\begin{algorithm}[H]
\caption{Randomized SVD (RSVD)}
\label{alg:rsvd}
\textbf{Function}: $\texttt{rsvd} \left( \textbf{A}, k \right) $
\begin{algorithmic}[1]
\STATE $\bm{\Omega} \gets \texttt{randn} \left( m, k \right) $  \hfill \texttt{// Generate random Gaussian matrix}
\STATE $\textbf{Y} \gets \textbf{A}  \bm{\Omega}$ \hfill \texttt{// Compute the sampled matrix Y}
\STATE $\left( \textbf{Q}, \textbf{R} \right) \gets \texttt{qr}\left( \textbf{Y}, 0 \right)$ \hfill \texttt{// Economy-sized QR decomposition to obtain orthonormal basis \textbf{Q}}
\STATE $\textbf{P} \gets \textbf{Q}^\top \textbf{A}$ \hfill \texttt{// Compute the small matrix \textbf{P}}
\STATE $(\tilde{\textbf{U}}, \textbf{S}, \textbf{V}) \gets \texttt{svd} \left( \textbf{P}, \texttt{'econ'} \right)$ \hfill \texttt{// Recover \textbf{S} and \textbf{V}}
\STATE $\textbf{U} \gets \textbf{Q}  \tilde{\textbf{U}}$   \hfill \texttt{// Recover \textbf{U}}
\end{algorithmic}
 \textbf{Return} $\left( \textbf{U}, \textbf{S}, \textbf{V} \right)$
\end{algorithm}

The selection of the rank $k$ in RSVD represents a delicate balance between approximation accuracy and dimensionality reduction. A higher rank yields a more accurate approximation of the original matrix $\mathbf{A}$ but retains a larger dimensionality. Conversely, a lower rank reduces dimensionality but may sacrifice accuracy. The challenge lies in identifying the optimal rank that balances these competing targets. 

In this work, RSVD is specifically adopted as a substitute for the classic SVD within the CoV-SSI algorithm described in Section~\ref{section21}. Referring to the notation introduced in the previous section, the Toeplitz matrix $\textbf{T}_{1|j_b}$ replaces matrix $A$ from Eq.~(\ref{eq:randomsamplingA}), which is sampled using matrix $\bm{\Omega}$. Therefore, the singular value decomposition of $\textbf{T}_{1|j_b}$ in Eq.~(\ref{Eq.Toeplitz}) is approximated by its low-rank representation via RSVD, as described by Eq.~(\ref{eq:reconstruct_A}). Given the model order limitations, the rank $k$ must be equal or larger than the maximum model order $N_{max}$ considered in the modal identification procedure. 

\subsection{3D Stabilization diagrams}\label{section23}

Among the various advantages of the CoV-SSI algorithm, such as its ease of automation, a notable limitation is that its results are not unique and are highly sensitive to the selection of numerous parameters. Additionally, identifying the entire structure can be particularly challenging due to issues like closely spaced modes and the dispersion of poles across different time lags and model orders. 
The application of RSVD in place of traditional SVD within the CoV-SSI modal identification alleviates the computational burden associated with extensive calculations. This offers new possibilities for conducting intensive pole stability analyses, with the potential of reducing the number of control parameters to be fixed by the analyst. In particular, the time lag parameter$j_b$ plays a key role in CoV-SSI. To avoid fixing this parameter, we propose conducting the modal identification across a range of time lags, which would be prohibitive with classical CoV-SSI. On this basis, the estimated poles are represented in a three-dimensional stabilization diagram, where the third axis is represented by the time-lag $\tau$. Consequently, the authors advocate for the implementation of RSVD not only to streamline computational efforts, but also to elevate the quality and performance of modal identification. The process of creating a 3D stabilization diagram encompasses the following steps:

\begin{enumerate}[label=\textit{\roman*}.]
\item \textit{Modal identification across different time-lag values.} It is necessary to select the values of $N_{min}$ and $N_{max}$ to define $\mathbb{M}$, as well as the minimum and maximum time-lag values to establish a domain $\mathbb{T}$ such that: $\tau \in \mathbb{T}, \mathbb{T} = [\tau_{\textrm{min}}, \tau_{\textrm{max}}] \in \mathbb{R}^+$. Subsequently, modal identification must be performed for each time-lag value defined in $\mathbb{T}$, ranging through the modal orders defined in $\mathbb{M}$.

\item \textit{3D stability analysis.} A strategy for removing spurious modes using SC and HC must be defined. For HC, spurious modes can be removed similarly to 2D stabilization diagram, as they are based on evaluation parameters related to a single pole. For SC, stability must be evaluated not only through contiguous model orders $m-1$ and $m$ but also through contiguous time-lags $t-1$ and $t$. Therefore, the proposed formulation sets a pole to be stable if it meets the following condition:
    \begin{equation}\label{eq:stabcriteria3D}
    \begin{aligned}
    &\frac{|f_i^m - f_j^{m-1}|}{\textrm{max}(f_i^m, f_j^{m-1})} & \vee & \quad \frac{|f_i^t - f_j^{t-1}|}{\textrm{max}(f_i^t, f_j^{t-1})} & \leq & \quad \alpha_f, \\
    &\frac{|\xi_i^m - \xi_j^{m-1}|}{\textrm{max}(\xi_i^m, \xi_j^{m-1})} & \vee & \quad \frac{|\xi_i^t - \xi_j^{t-1}|}{\textrm{max}(\xi_i^t, \xi_j^{t-1})} & \leq & \quad \alpha_{\xi},  \\
    &\textrm{MAC}\left(\bm{\phi}^{m}_i, \bm{\phi}^{m-1}_j\right) & \vee & \quad \textrm{MAC}\left(\bm{\phi}^{t}_i, \bm{\phi}^{t-1}_j\right) & \leq & \quad \alpha_{\textrm{MAC}}, 
    \end{aligned}
    \end{equation}
    where: $i$ and $j$ are different poles, and $\alpha_f$, $\alpha_\xi$, and $\alpha_{\textrm{MAC}}$ are the thresholds in terms of frequency, damping ratio, and MAC, respectively.

\item \textit{3D clustering.} A two-stage clustering approach is adopted, similar to the method proposed by Charbonnel~\cite{Charbonnel2021}, where a fuzzy C-means clustering algorithm is followed by a hierarchical clustering one. The use of fuzzy clustering (Stage I) aims to further investigate the stability of the previously retained poles. In this stage, the previously identified stable poles are collected and divided into two clusters (possibly physical and certainly unphysical) based on a comprehensive evaluation of their stability (along the time lag and model order axes) in terms of frequency, damping ratio, MAC, MPC, and MPD values. Only the resulting cluster of possibly physical poles are retained for Stage II. This stage involves a hierarchical clustering algorithm that constructs a hierarchical $\Pi$-shaped dendrogram based on a distance metric defined as the sum of the relative differences of the retained poles in terms of frequencies, damping and MAC. Finally, the physical modes of the structure, derived from the 3D stabilization diagram, can be identified by imposing a threshold for the distance metric and a minimum cluster size.
\end{enumerate}

The 3D stabilization diagram mitigates these challenges by enabling partial automation of the CoV-SSI identification process. Furthermore, using the 3D stabilization diagram for continuous modal identification allows for the extraction of more accurate and stable results as demonstrated in the following results and discussion section. This improvement is achieved by considering a broader range of control parameters, including variations in the time lag, which aids in refining and stabilizing the results.

\begin{figure}[H]
\centering
   \includegraphics[width=1\textwidth]{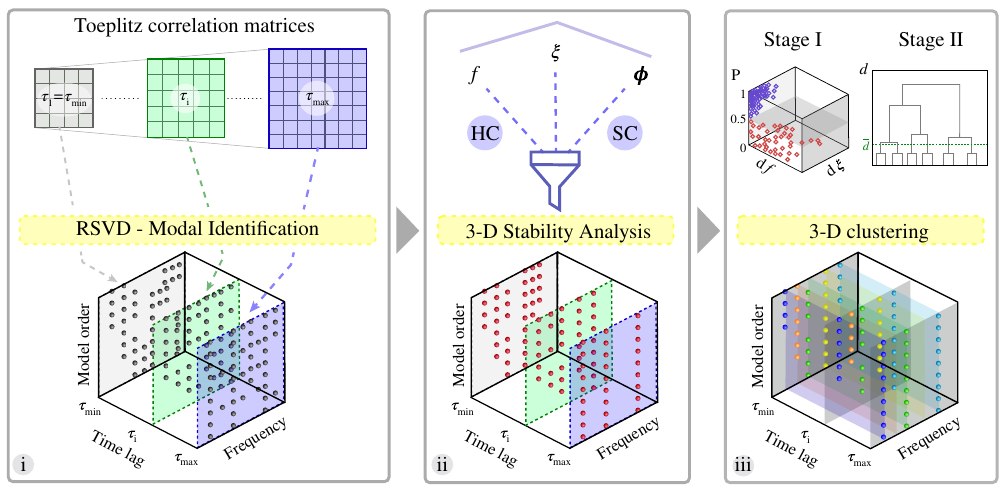}
   \caption{Flowchart for the construction of 3D stabilization diagrams considering pole stability across various model orders and time lag values.}    
    \label{fig:flowchart}
\end{figure}


\section{Numerical results and discussion}\label{section3}

In this section, the methodologies proposed in Section~\ref{section2} are applied to a theoretical system with 10-DOFs and to a real-world case study: the San Faustino Bridge in Italy. The objectives of this discussion include: (i) to delineate empirical guidelines for the effective selection of key parameters within the CoV-SSI algorithm, (ii) to establish a rigorous methodology for determining the minimum rank required when employing RSVD, taking into account the varying sizes of the Toeplitz matrix, (iii) to quantify the computational savings achieved by using RSVD compared to traditional SVD, and (iv) to evaluate the long-term performance of the proposed 3D stabilization diagrams in comparison to 2D stabilization diagrams.

\subsection{Theoretical 10-DOFs case study}\label{section31}

\begin{figure}[b]
\centering
   \includegraphics[width=1\textwidth]{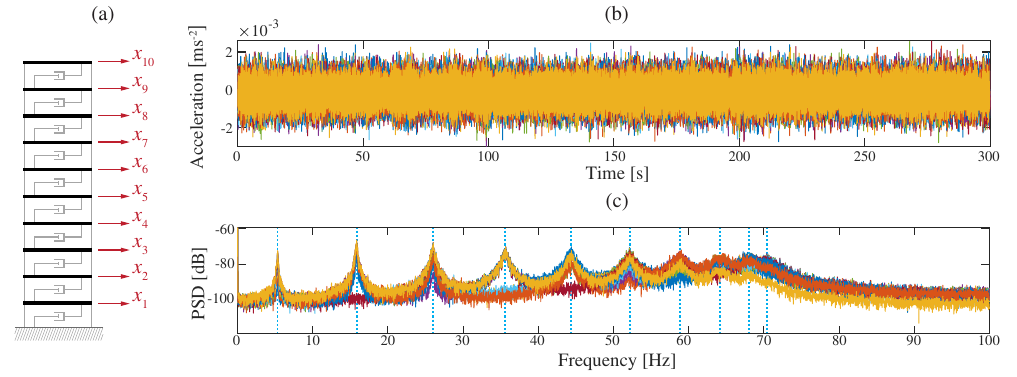}
   \caption{Sketch of the 10-DOFs dynamic system (a), sampled accelerations with signal-to-noise ratio (SNR) of 20 dB (b), and corresponding power spectral density (PSD) spectrum (c).}    
    \label{fig:10dofsystem}
\end{figure}

Let us consider the 10-DOFs theoretical dynamic system illustrated in Fig.\ref{fig:10dofsystem}a. This system is modeled as a damped shear-type 10-storey frame, where the structural mass \textbf{M} and stiffness \textbf{K} matrices are constructed based on each floor having a mass and a stiffness, set at 100 kg and 5000 kN/m, respectively. The damping matrix \textbf{C} is defined proportionally to the mass matrix \textbf{M} (diagonal) and the stiffness matrix \textbf{K} (positive-definite) using the Rayleigh method, imposing that the first and fourth angular frequencies of the system have a damping ratio $\xi$ of 1\%. The natural frequencies, obtained from the eigenvalue decomposition of the system matrices, along with their corresponding numerical damping ratios, are detailed in Table \ref{Tab1}.

\begin{table}[H]
\setlength{\tabcolsep}{3pt} 
\newcommand\Tstrut{\rule{0pt}{0,3cm}}         
\newcommand\Bstrut{\rule[-0.15cm]{0pt}{0pt}}   
 \footnotesize		
 \caption{Theoretical natural frequencies $f_i$ and damping ratios $\xi_i$ of the 10-DOFs system.}
 \vspace{0.1cm}
   \centering
   \begin{tabular}{lllp{0.5cm}lll}
   \hline
   
   Mode & $f_i$ [Hz] & $\xi_i$ [\%] &  & Mode & $f_i$ [Hz] & $\xi_i$ [\%] \Tstrut\Bstrut\\
   \hline
    1 & 5.319 & 1.000 & & 6 & 52.176 & 1.364 \Tstrut\\
    2 & 15.838 & 0.679 &  & 7 & 58.809 & 1.516 \Tstrut\\
    3 & 26.004 & 0.814 & &8 & 64.128 & 1.640 \Tstrut\\
    4 & 35.588 & 1.000 &  &9 & 68.014 & 1.731 \Tstrut\\
    5 & 44.378 & 1.189 &  & 10 & 70.381 & 1.786 \Tstrut\\
   \hline
   \end{tabular}
   \label{Tab1}
\end{table}

\subsubsection{Parametric analysis: rank selection}\label{section312}

The main user-defined parameter for computing the RSVD is the rank $k$. Therefore, substituting SVD with the RSVD algorithm, as proposed in Section~\ref{section21}, necessitates calibrating the minimum rank required for accurate results, while keeping the other CoV-SSI identification parameters unchanged. A parametric analysis was performed using signals derived from the discrete-time state-space model of the 10-DOFs system. The matrices $\textbf{A}$ and $\textbf{C}$ in Eq.~(\ref{Eq.DSSM}) were used to generate the observation matrix $\textbf{Y}$. Noise was simulated as a zero-mean Gaussian process with signal-to-noise ratio (SNR) of 10 dB, 15 dB, 20 dB, and 25 dB, sampled at 200 Hz. Four observation matrices were computed, each based on 5-minute-long acceleration time histories differing only in noise contribution. The 10 time histories and the corresponding power spectral density (PSD) curves of the signals with SNR=20 dB are illustrated in Figs.~\ref{fig:10dofsystem}(a) and~\ref{fig:10dofsystem}(b).
The primary objective was to determine whether the minimum rank $k$ required for accurate results is influenced by noise and to establish its value as the size of the Toeplitz matrix varies. To achieve this, a parametric analysis was conducted by iterating the proposed RSVD-CoV-SSI method. This analysis involved incrementally increasing the rank $k$ and comparing the results with those derived from the classical CoV-SSI algorithm using SVD. The iteration continued until the modal characteristics provided by the algorithm using SVD converged with those obtained using RSVD for different sizes of the Toeplitz matrix. Each simulation involved varying the time-lag step $j_b$ from 100 to 1000 in increments of 100, resulting in Toeplitz matrices with sizes ranging from 1000 to 10000 and corresponding to time-lags $\tau$ between 1 and 10 seconds, as determined by the following rule:

\begin{equation}\label{eq:tlag}
    \tau = \frac{2 j_b - 1}{f_s}.
\end{equation}

Generally, the required model order for obtaining reliable results depends on both the number of channels (and hence the complexity of the sensor setup) and the level of noise in the data acquisition. To consider suitable model orders for each set of time histories corresponding to different SNR values, model orders ranging from 10 to 100, with increments of 2, were examined. The HC applied in the computation of the stabilization diagrams included discarding poles with damping ratios $\xi_i > 10\%$, MPC$\leq 60\%$, and MPD$\geq 50\%$. Additionally, the SC involved discarding consecutive modes if the relative differences in frequencies $f$, damping ratios $\xi$, and MAC values between different model orders exceeded 2\%, 2\%, and 5\%, respectively. To compare and match modal characteristics, the maximum allowable relative differences in frequencies, damping ratios, and MAC values were set to 1\%, 1\%, and 2\%, respectively.

It is important to remark that the selected wide range of increasing time lags has the sole purpose of determining the minimum rank needed for RSVD to accurately replicate the same obtained through SVD, regardless of the quality of the subsequent modal identification. Nevertheless, when applied for modal identification, excessively high or low time lag values can significantly compromise the quality of the model identification. For this reason, some guidelines for the range of time lags to be considered in the 3D stabilization diagram are introduced hereafter in Section~\ref{section313}.

\begin{figure}[H]
\centering
   \includegraphics[scale=1]{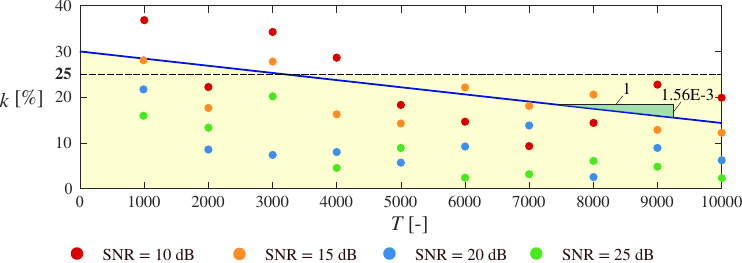}
   \caption{Results of the parametric analysis for the 10-DOFs dynamic system. The dots represent the experimental minimum percentage ranks $\overline{k}$ for different SNR values.}    
    \label{fig:parametric_analysis}
\end{figure}

The results of the parametric analyses are presented in Fig.~\ref{fig:parametric_analysis}, where the rank $k$ is expressed as a percentage of the Toeplitz matrix size $T = j_b \cdot l$. Denoting with $\overline{k}$ the minimum percentage rank required to ensure accurate results, it can be observed that its value decreases as the size of the Toeplitz matrix increases. This decreasing trend in the minimum rank $\overline{k}$ as the size of the Toeplitz matrix $T$ increases suggests that excessively large time lags do not add informational content but instead introduce additional noise, allowing lower percentage ranks to capture the necessary matrix information effectively. Conversely, higher percentage ranks are required to retain all essential information in smaller matrices. Furthermore, for signals with high noise levels (SNR of 10 dB or 15 dB), $\overline{k}$ tends to be higher compared to those with low noise (SNR of 20 dB or 25 dB). Notably, for signals with low noise, $\overline{k}$ consistently remains below 25\%. Therefore, for signals with high SNR values, a default choice of $\overline{k} = 25\%$ might be effective regardless of $T$. To develop an empirical rule for guiding the selection of $\overline{k}$ as a function of $T$, a line interpolating all points in the graph was computed. The intercept on the $\overline{k}$ axis was then adjusted to 30\% by slightly vertically shifting the interpolating line to highlight that, generally, the minimum rank falls between 25\% and 30\%. Ultimately, the empirical formula to determine $\overline{k}$ is:

\begin{equation} \label{eq:empirical_formula}
    \overline{k} \, \left(T \right) = \textrm{max} \left\{  30 - 0.00156 \, T   \, ; \, 25 \right\}.
\end{equation}

This formula, depicted by the blue dashed line in Fig.~\ref{fig:parametric_analysis}, provides a practical method for selecting the rank percentage based on the size of the Toeplitz matrix, with the equation graphically represented by the blue line in Fig.~\ref{fig:parametric_analysis}. It can be noted that some points related to SNR values of 10 dB and 15 dB lie above the function expressed in Eq.~(\ref{eq:empirical_formula}). Nevertheless, these points were disregarded because, in practical applications, the signals acquired for SHM purposes typically exhibit higher SNR values. It is important to note that the empirically derived formulation in Eq.~(\ref{eq:empirical_formula}) for the minimum rank, determined from this theoretical case study, serves as initial guidance, with the understanding that further calibration may be needed for specific applications.

\subsubsection{2D and 3D stabilization diagrams including the RSVD}\label{section313}

\begin{figure}[t]
\centering
   \includegraphics[width=1\textwidth]{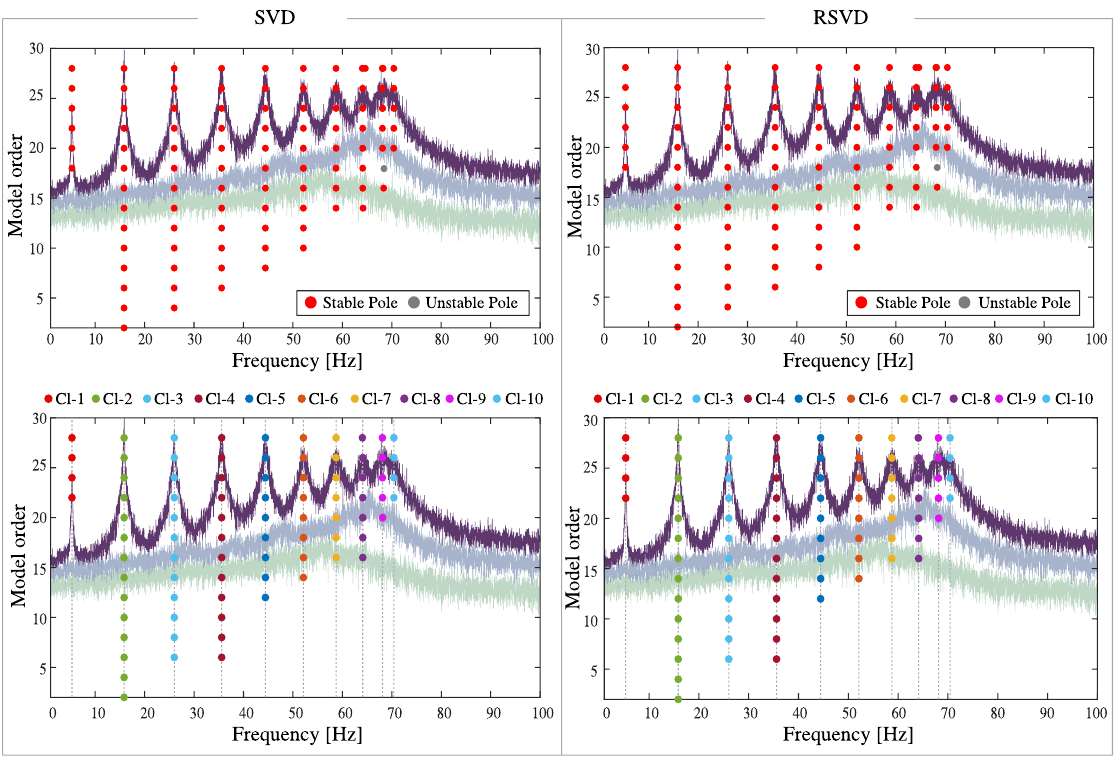}
   \caption{Stabilization diagrams extracted by CoV-SSI using SVD and RSVD. In the background, the first three singular values of the spectral density matrix are plotted for reference.}    
    \label{fig:RSVD_SVD_10dof}
\end{figure}

Considering the results collected from the parametric analysis reported in Section~\ref{section312}, a comparison of the performance of SVD and RSVD for extracting both classical 2D and 3D stabilization diagrams was conducted. For this purpose, the signals with an SNR value of 20 dB were used. Specifically, due to the low noise level considered in this case, modal identifications were performed using model orders ranging from 2 to 30 with increments of 2. The time-lag $\tau$ was selected based on the minimum time-lag step $j_b$ as specified by the empirical formulation proposed by Reynders \textit{et al.} in~\cite{reynders2008reference}, which is given by:

\begin{equation}\label{eq:jb}
    j_b = \textrm{int} \left( 10 \frac{f_s}{2f_0} \right),
\end{equation}

\noindent where $f_0$ is the fundamental natural frequency of the structural system.Therefore, $j_b$ is equal to 189, resulting in a corresponding time-lag $\tau$ of 1.885 seconds, which forms a Toeplitz matrix of size 1890$\times$1890. The rank $r$ used for the computation of the RSVD was defined as the minimum rank according to Eq.~(\ref{eq:empirical_formula}), which amounts to 27.05\%. Consequently, 512 columns out of 1890 were randomly selected from the initial Toeplitz matrix to compute the RSVD. The comparison between the stabilization diagrams derived from SVD and RSVD is illustrated in Fig.~\ref{fig:RSVD_SVD_10dof}. The same HC and SC specified in Section~\ref{section312} were applied to eliminate spurious poles, and the identical clustering strategy was employed. It can be observed that the stabilization diagrams are identical, indicating that RSVD provides results comparable to those obtained using traditional SVD. Additionally, the computational time for performing the SVD is 1.77 seconds, whereas using the RSVD reduces this time to 0.32 seconds, making the analysis much more efficient.

\begin{figure}[t]
\centering
   \includegraphics[width=1\textwidth]{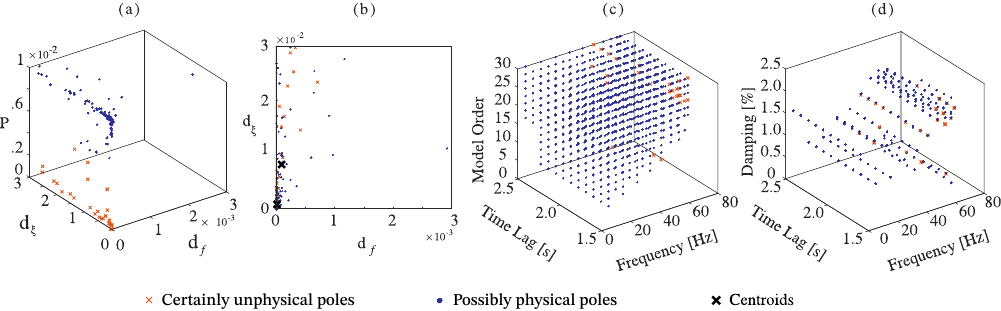}
   \caption{Stability analysis of the 10DOFs system after the first stage of clustering.}    
    \label{fig:3d_poles_10dof}
\end{figure}

The 3D stabilization diagram was created following the methodology outlined in Section~\ref{section23}. The initial step in this approach involves selecting the minimum and maximum time lags to construct a 3D space that is adaptable to the characteristics of the structural system. To achieve this, a variation of the rule defined in Eq.~(\ref{eq:jb}) is employed:

\begin{equation}\label{eq:jb_3D}
\begin{aligned}
    & j_{b,\textrm{min}} = \textrm{int} \left( 9 \frac{f_s}{2f_0} \right), \\
    & j_{b,\textrm{max}} = \beta j_{b,\textrm{min}},
\end{aligned}
\end{equation}

\noindent where $\beta$ is a multiplier reasonably chosen to be 1.5. Given that the range between $j_{b,\textrm{min}}$ and $j_{b,\textrm{max}}$ is divided into 10 intermediate steps, spanning from 170 to 251, the time lags according to Eq.~(\ref{eq:tlag}) define a domain between 1.70 seconds and 2.55 seconds. After performing the modal identification for each value of the time lag, stability analyses of the poles were conducted by evaluating each pair of poles along both the time lag and model order axes. In accordance with the criteria defined in Eq.~(\ref{eq:stabcriteria3D}), the SC thresholds $\alpha_f$, $\alpha_\xi$, and $\alpha_{\textrm{MAC}}$ were set to 1\%, 3\%, and 2\%, respectively. The same HC applied to the 2D cases was used. Subsequently, the two-stage clustering process was applied, considering the stable poles across the entire 3D space. The results of the first stage are presented in Fig.~\ref{fig:3d_poles_10dof}. In particular, Fig.~\ref{fig:3d_poles_10dof}a shows the distribution in 3D of the previously defined stable poles according to their distance metrics $d_f$ and $d_\xi$, defined as:

\begin{equation}
    d_f^{ij} = \frac{|f_i - f_j|}{\textrm{max}(f_i, f_j)}, \hspace{0.5cm} d_\xi^{ij} = \frac{|\xi_i - \xi_j|}{\textrm{max}(\xi_i, \xi_j)},
\end{equation}

\begin{figure}[H]
\centering
   \includegraphics[width=1\textwidth]{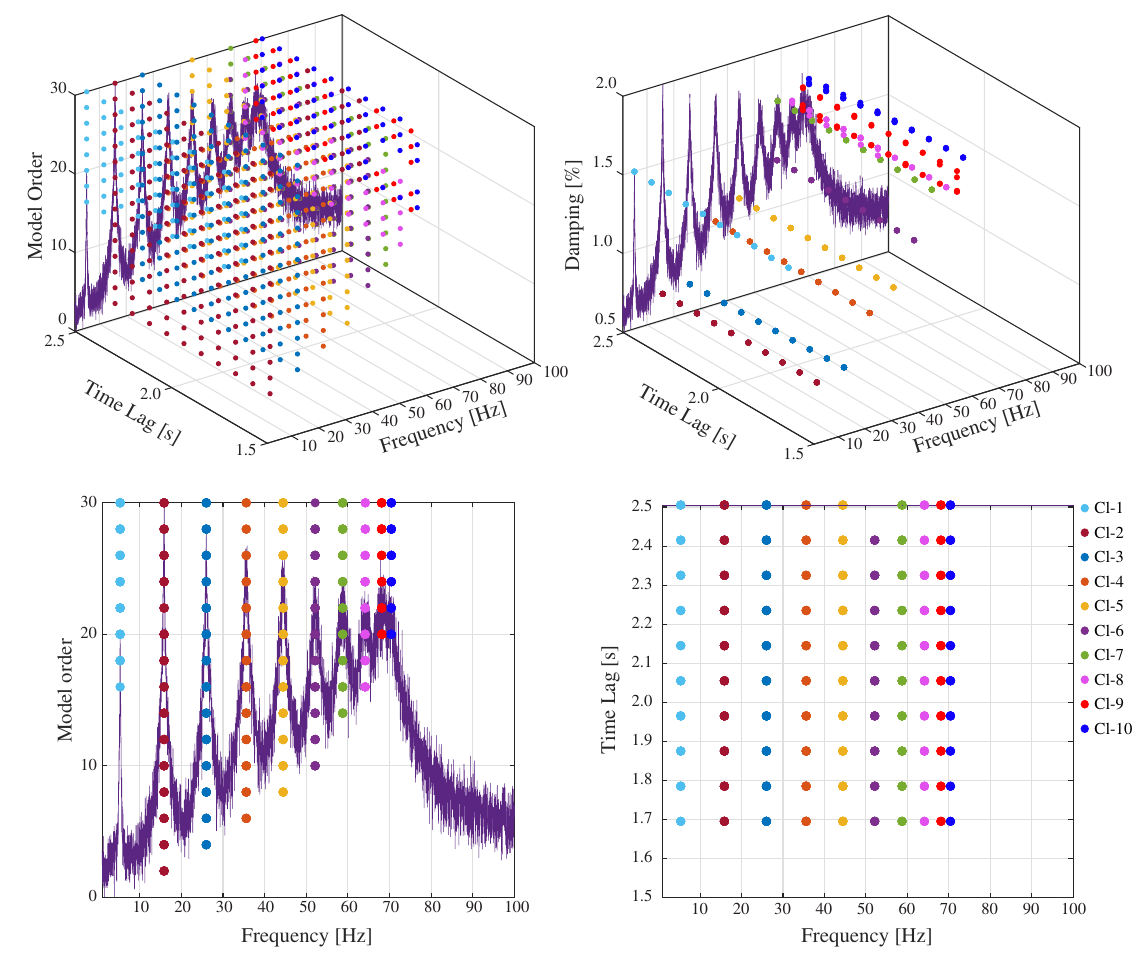}
   \caption{Modal identification results of the 10-DOFs dynamic system by 3D stabilization diagrams using RSVD-CoV-SSI. In the background (on the frequency-model order plane), the first singular value of the spectral density matrix is plotted for reference.}    
    \label{fig:clusters_10dof}
\end{figure}

\noindent where $f_i$, $f_j$, $\xi_i$, $\xi_j$ are the $i$-th and $j$-th frequencies and damping ratios in the 3D space. The fuzzy C-means clustering algorithm is adopted to further refine the stabilization diagram by partitioning the poles into two clusters: possibly physical and certainly unphysical poles. To this end, the algorithm considers a distance metric involving relative differences in frequency, damping ratio, MPC, and MPD, as well as 1-MAC values. Based on this metric, a partitioning matrix is iteratively generated, assigning a fuzzy membership grade $P \in [0,1]$ to each pole, indicating the degree to which each pole belongs to each of the two clusters (refer to references~\cite{Bezdek1981fuzzycmean,Babuska2000fuzzycmean} for further details). The fuzzy membership grade $P$ allows the poles to be divided into the cluster of possibly physical poles if $P \geq 0.5$, or into the cluster of certainly unphysical poles if $P < 0.5$. Consequently, only the poles classified as possibly physical proceed to the second stage of clustering. Figure~\ref{fig:3d_poles_10dof}b presents a 2D view of the clusters, including the centroids of the possibly physical and certainly unphysical clusters. Note in this figure that the centroid of the cluster of certainly unphysical poles is located far from the origin, while the centroid of the cluster of possibly physical poles is positioned very close to the origin. Figures~\ref{fig:3d_poles_10dof}c and~\ref{fig:3d_poles_10dof}d display the 3D stabilization diagrams in the Frequencies $f$ - Time lag $\tau$ - Model order $N$ space and the Frequencies $f$ - Time lag $\tau$ - Damping ratios $\xi$ space, respectively. Finally, the results of the 3D clustering are shown in Fig.~\ref{fig:clusters_10dof}.

The comprehensive comparison of the results is reported in Table~\ref{Tab2}, where the percentile parameters used to compare frequencies, damping ratios, and mode shapes are defined as follows:

\begin{equation}
\begin{aligned}
   & \Delta_{f,\textrm{RSVD}} = \frac{|f_{\textrm{SVD}}-f_{\textrm{RSVD}}|}{f_{\textrm{SVD}}}\cdot 100, && \Delta_{\xi,\textrm{RSVD}} = \frac{|\xi_{\textrm{SVD}}-\xi_{\textrm{RSVD}}|}{\xi_{\textrm{SVD}}}\cdot 100, && \textrm{MAC}_{\textrm{RSVD}}=\textrm{MAC} \left( \bm{\phi}_{\textrm{SVD}},\bm{\phi}_{\textrm{RSVD}} \right), \\
   & \Delta_{f,\textrm{3D}} = \frac{|f_{\textrm{SVD}}-f_{\textrm{3D}}|}{f_{\textrm{SVD}}}\cdot 100, && \Delta_{\xi,\textrm{3D}} = \frac{|\xi_{\textrm{SVD}}-\xi_{\textrm{3D}}|}{\xi_{\textrm{SVD}}}\cdot 100, && \textrm{MAC}_{\textrm{3D}}=\textrm{MAC} \left( \bm{\phi}_{\textrm{SVD}},\bm{\phi}_{\textrm{3D}} \right), \\
\end{aligned}
\end{equation}

\noindent where $\Delta_{f,\textrm{RSVD}}$ and $\Delta_{f,\textrm{3D}}$ represent the relative differences in frequencies between those derived from the reference classic modal identification involving the SVD ($f_{\textrm{SVD}}$) and those evaluated considering the RSVD in 2D ($f_{\textrm{RSVD}}$) or in 3D ($f_{\textrm{3D}}$). Similarly, $\Delta_{\xi,\textrm{RSVD}}$ and $\Delta_{\xi,\textrm{3D}}$ compare the damping ratios derived from the SVD ($\xi_{\textrm{SVD}}$) with those from the RSVD in 2D ($\xi_{\textrm{RSVD}}$) and 3D ($\xi_{\textrm{3D}}$). The comparison in terms of MAC \cite{allemang2012MAC} between the mode shapes identified using the SVD ($\bm{\phi}_{\textrm{SVD}}$) and those identified using the RSVD through the 2D ($\bm{\phi}_{\textrm{RSVD}}$) and 3D ($\bm{\phi}_{\textrm{3D}}$) stabilization diagrams are represented by $\textrm{MAC}_{\textrm{RSVD}}$ and $\textrm{MAC}_{\textrm{3D}}$, respectively.

The comprehensive results of the analyses involving the previously defined parameters are reported in Table~\ref{Tab2}. Notably, the accuracy in terms of frequencies is very high, with a maximum relative difference of 0.004\% in the case of RSVD-CoV-SSI and 0.025\% when the RSVD is used to compute the 3D stabilization diagram. On the other hand, slightly larger differences are observed in damping ratios, with a maximum relative difference of 0.499\% and 1.943\%, respectively. However, it is commonly known in the literature that damping ratios derived from CoV-SSI tend to exhibit more variability compared to frequencies (see e.g.~\cite{lorenzoni2019ambient}); therefore, the larger errors observed in damping ratios do not compromise the validity of the results. Additionally, the accuracy in terms of mode shapes is exceptionally high for both the RSVD-CoV-SSI and RSVD in 3D stabilization diagrams, achieving a perfect match for each mode shape, as indicated by $\textrm{MAC}_{\textrm{RSVD}}$ and $\textrm{MAC}_{\textrm{3D}}$ values of 1 for each mode. It can also be noted that the results from the 3D stabilization diagram tend to present larger differences compared to the sole use of RSVD. However, this should not be generalized as an error, because these modal features result from a more comprehensive and extensive analysis. Therefore, this approach can only improve the reliability of the results compared to the standard method of computing stabilization diagrams. 

\begin{figure}[H]
\centering
   \includegraphics[scale=1.0]{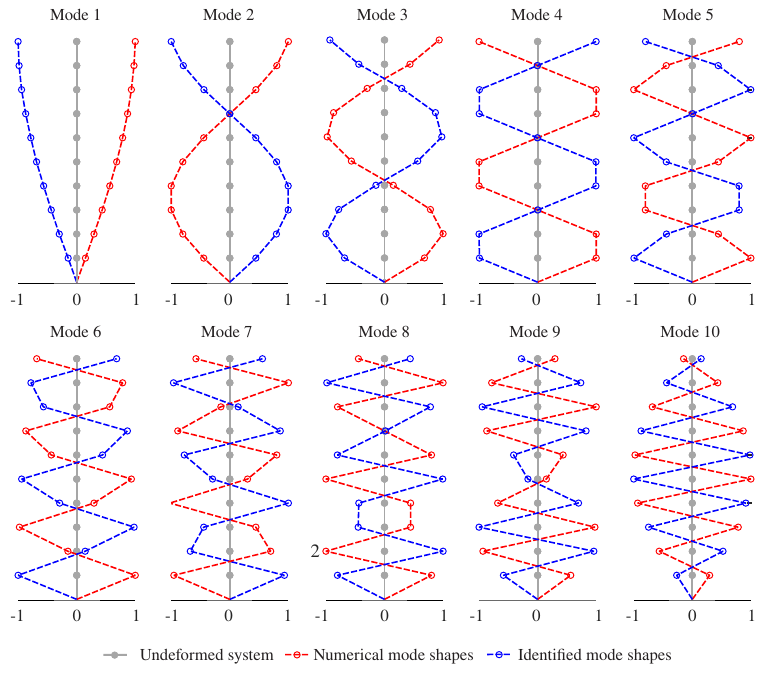}
   \caption{Numerical and identified mode shapes of the 10DOFs system.}    
    \label{fig:10dofmodeshapes}
\end{figure}

In Fig.~\ref{fig:10dofmodeshapes}, the identified and numerical mode shapes of the 10-DOFs dynamic system are presented.

\begin{table}[H]
\setlength{\tabcolsep}{3pt} 
\newcommand\Tstrut{\rule{0pt}{0,3cm}}         
\newcommand\Bstrut{\rule[-0.15cm]{0pt}{0pt}}   
 \footnotesize		
 \caption{Comparison between the modal identification results derived from the 2D stabilization diagram by using SVD ($f_{\textrm{SVD}}$, $\xi_{\textrm{SVD}}$) and the RSVD ($f_{\textrm{RSVD}}$, $\xi_{\textrm{RSVD}}$) and 3D stabilization diagram ($f_{\textrm{3D}}$, $\xi_{\textrm{3D}}$).}
 \vspace{0.1cm}
   \centering
   \begin{tabular}{cp{0.5cm}cccccp{0.5cm}cccccp{0.5cm}cc}
   \hline	
   \multirow{2}{*}{Mode} & & \multicolumn{5}{c}{Frequencies [Hz]} & & \multicolumn{5}{c}{Damping ratios  [\%]} & & \multicolumn{2}{c}{Mode shapes}\Tstrut\Bstrut\\
   \cline{3-16}
  &  & $f_{\textrm{SVD}}$ & $f_{\textrm{RSVD}}$ & $f_{\textrm{3D}}$ & $\Delta_{f,\textrm{RSVD}}$  & $\Delta_{f,\textrm{3D}}$ & & $\xi_{\textrm{SVD}}$ & $\xi_{\textrm{RSVD}}$ & $\xi_{\textrm{3D}}$ &  $\Delta_{\xi,\textrm{RSVD}}$ & $\Delta_{\xi,\textrm{3D}}$ & & MAC$_{\textrm{RSVD}}$ & MAC$_{\textrm{3D}}$  \Tstrut\Bstrut\\
       \hline
1 & & 5.312 & 5.312 & 5.313 & \textbf{0.003} & 0.020 & & 1.465 & 1.472 & 1.480 & \textbf{0.499} & 1.053 &  & 1.000 & 1.000 \Tstrut\\
2 & & 15.856 & 15.856 & 15.858 & 7.9E-05 & 0.009 & & 0.677 & 0.677 & 0.673 & 0.012 & 0.485 &  & 1.000 & 1.000 \Tstrut\\
3 & & 26.002 & 26.002 & 26.002 & 1.5E-04 & 0.001 & &  0.707 & 0.707 & 0.696 & 0.051 & 1.517 &  & 1.000 & 1.000 \Tstrut\\
4 & & 35.589 & 35.589 & 35.590 & 2.8E-05 & 0.003 & &  1.021 & 1.020 & 1.022 & 0.038 & 0.151 &  & 1.000 & 1.000 \Tstrut\\
5 & & 44.438 & 44.438 & 44.438 & 7.1E-05 & 9.3E-05 & & 1.131 & 1.131 & 1.130 & 0.001 & 0.140 &  & 1.000 & 1.000 \Tstrut\\
6 & & 52.121 & 52.121 & 52.121 & 9.3E-04 & 9.7E-04 & &  1.393 & 1.392 & 1.393 & 0.037 & 0.030 &  & 1.000 & 1.000 \Tstrut\\
7 & & 58.724 & 58.723 & 58.724 & 3.4E-04 & 0.001 & & 1.679 & 1.680 & 1.677 & 0.008 & 0.148 &  & 1.000 & 1.000 \Tstrut\\
8 & & 64.116 & 64.115 & 64.131 & 4.6E-04 & \textbf{0.025} & & 1.621 & 1.621 & 1.653 & 0.036 & \textbf{1.943} &  & 1.000 & 1.000 \Tstrut\\
9 & & 68.112 & 68.113 & 68.105 & 0.002 & 0.011 & & 1.661 & 1.662 & 1.651 & 0.077 & 0.590 &  & 1.000 & 1.000 \Tstrut\\
10 & & 70.415 & 70.415 & 70.411 & 9.9E-05 & 0.005 & & 1.813 & 1.814 & 1.783 & 0.064 & 1.623 &  & 1.000 & 1.000 \Tstrut\\

   \hline
   \end{tabular}
   \label{Tab2}
\end{table}

\subsection{Real case study: the San Faustino Bridge}\label{section32}

\begin{figure}[H]
\centering
   \includegraphics[width=1\textwidth]{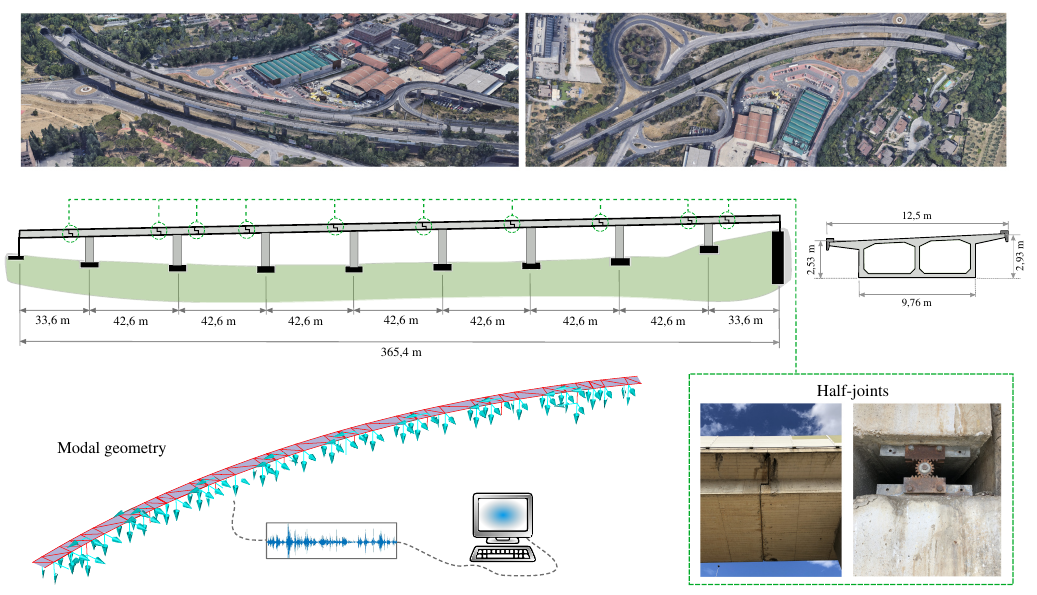}
   \caption{Geometry of the San Faustino bridge and modal geometry of the acceleration channels.}    
    \label{fig:sanfaustino}
\end{figure}

The San Faustino Bridge is a nine-span post-tensioned concrete box-girder bridge located in Perugia, in the Italian region of Umbria (Fig.~\ref{fig:sanfaustino}). It is one of two parallel twin bridges serving the same stretch of road, both with an identical static scheme. Specifically, the bridge under investigation carries traffic in the direction to Rome, while the twin bridge carries traffic towards Florence. The total length of the bridge is 365.4 m. The first and ninth spans are 33.6 m long, and the central spans are 42.6 m long. There are 10 half-joints on the bridge: one for each span, except for the third span which has two. The monitoring system comprises 66 MEMS accelerometers (±2 g, 24-bit ADC, noise density 22.5 $\mu g/\sqrt{\text{Hz}}$) distributed along the entire structure, resulting in 114 measurement channels. Each span is equipped with two biaxial accelerometers in the $y$- (transversal) and $z$-directions (vertical), and two uniaxial sensors in the $z$-direction (vertical). Each pile is equipped with a biaxial accelerometer on the top; the first and last piles are also equipped with a triaxial accelerometer at the base. Additionally, the half-joints are equipped with a triaxial accelerometer and a uniaxial accelerometer in the $z$-direction. The monitoring system is connected to an onboard edge computer via a wired connection, continuously storing 12 acceleration files per day, each with a duration of two hours and a sampling acquisition frequency of 125 Hz. The acceleration records are processed through a simple filtering sequence involving the elimination of linear trends and downsampling to 40 Hz (eighth-order low-pass Chebyshev Type I filter with a cut-off frequency of $0.8 \cdot 20$ Hz followed by a decimation to 40 Hz).

\subsubsection{Computational efficiency}\label{section321}

\begin{figure}[t]
\centering
   \includegraphics[scale=1.0]{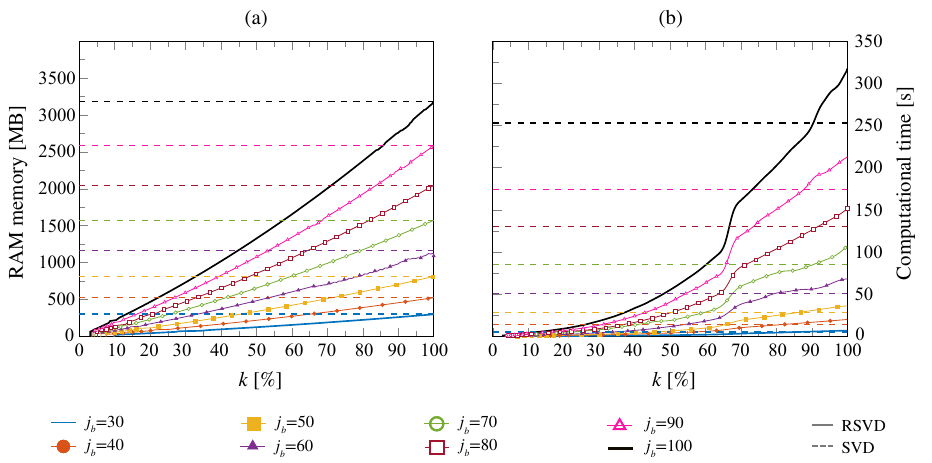}
   \caption{Comparison between the computational burdens in terms of RAM memory (a) and time (b) of SVD and RSVD as the rank $k$ increases.}    
    \label{fig:computational_burden}
\end{figure}

The data from the previously presented bridge was utilized in the following analysis to (i) perform a deep investigation of the computational performance of RSVD compared to SVD, and (ii) to define the experimental minimum percentile rank $\overline{k}$ to validate Eq.~(\ref{eq:empirical_formula}) on a real case study. To this end, the filtered acceleration time series recorded on April 25\textsuperscript{th}, 2023, at 2 p.m. were analyzed by considering different time-lags to generate Toeplitz matrices of varying sizes. The computational burden associated with computing the SVD of each matrix was compared with that required to perform the RSVD, considering different percentage ranks $k$ uniformly varied between 0 and 100\%. As $k$ increased, the identified modal characteristics evaluated using the RSVD were compared to those obtained via SVD to determine the experimental $\overline{k}$ for the bridge. This comparison was conducted using the same parameters defined in Section~\ref{section312}. Toeplitz matrices were generated with time lag steps $j_b$ uniformly sampled between 30 and 80 with a step of 10, resulting in matrix sizes ranging from 3420 to 11400. The comprehensive results of the analysis are depicted in Fig.~\ref{fig:computational_burden}. Specifically, Fig.~\ref{fig:computational_burden}a illustrates the computational burden in terms of RAM memory required to compute the SVD or the RSVD, while Fig.~\ref{fig:computational_burden}b shows the computational time involved in the analysis. Each curve in this figure represents the computational burden associated with increasing the percentage rank $k$; the dashed lines indicate the computational memory and time required for computing the SVD of the entire matrix. All computational burdens refer to hardware equipped with an Intel\textsuperscript{\textregistered} Core\textsuperscript{TM} i7-8750H CPU at 2.2 GHz and 16 GB of installed RAM.

\begin{figure}[b]
\centering
   \includegraphics[scale=1]{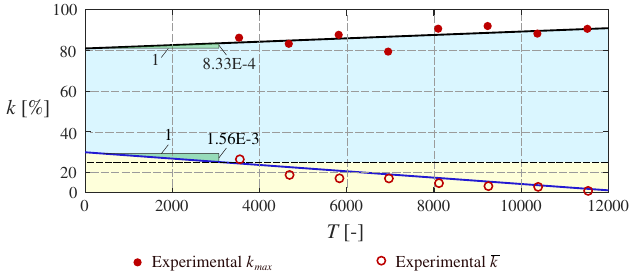}
   \caption{Parametric analysis of the San Faustino bridge. Filled dots represent the experimental maximum percentage rank  $k_{max}$ at which the computational time for computing RSVD equals that of computing SVD. Open circles indicate the experimental minimum percentage rank $\overline{k}$.}
    \label{fig:Parametric_SF}
\end{figure}

It can be observed that, as expected, the computational burden associated with performing the RSVD increases with the percentage rank $k$. This trend is not linear as it rises and tends to rise more sharply as the size of the Toeplitz matrix $T$ increases. As illustrated in Fig.~\ref{fig:computational_burden}a, the computational memory required to perform the SVD of each of the analyzed matrices is equivalent to that involved in computing the RSVD of the complete matrix, i.e., when $k$ equals 100\%. However, as shown in Fig.~\ref{fig:computational_burden}b, there exists a value of $k$ above which the computational time required for the RSVD exceeds that required for the SVD. This observation is expected, as the RSVD involves additional intermediate steps to compute the reduced matrix and its QR decomposition before applying the SVD. Consequently, if the size of the matrix on which the factorization is computed becomes too large, the total time required for the RSVD may surpass that needed for the SVD. Denoting by $k_{max}$ the percentage rank at which the computational time for the RSVD and the SVD are equal, it is evident from Fig.~\ref{fig:Parametric_SF} that $k_{max}$ tends to increase linearly with the size of the Toeplitz matrix $T$. The black line in this figure represents the linear interpolation of the experimental $k_{max}$ values for the San Faustino bridge, indicated by the filled points. The dashed black line and the blue line correspond to the minimum percentage rank $\overline{k}$ as derived from Eq.~(\ref{eq:empirical_formula}). It can be observed that the experimental $\overline{k}$ values from the San Faustino analysis align well with the empirical rule, thereby validating the proposed formula. The shaded blue area between the black and blue lines encompasses the range of $k$ values that allow for computational time and memory savings while ensuring accurate results when performing the RSVD. Furthermore, Fig.~\ref{fig:computational_burden} demonstrates that selecting the minimum rank $\overline{k}$ from Eq.~(\ref{eq:empirical_formula}) can yield substantial computational savings. For instance, with a Toeplitz matrix size $T$ of 11400 ($j_b = 100$), the computational time required for the SVD is 317.36 s, while performing the RSVD with a rank $r$ of 25\% takes only 13.04 s, resulting in a 95.89\% reduction in computational time. In terms of memory, the RSVD requires 596.66 MB compared to 3171.27 MB for the SVD, which represents an 81.19\% reduction, effectively reducing the memory requirement from GB to MB.

\subsubsection{Continuous RSVD-CoV-SSI modal identification}\label{section322}

\begin{figure}[H]
\centering
   \includegraphics[width=1\textwidth]{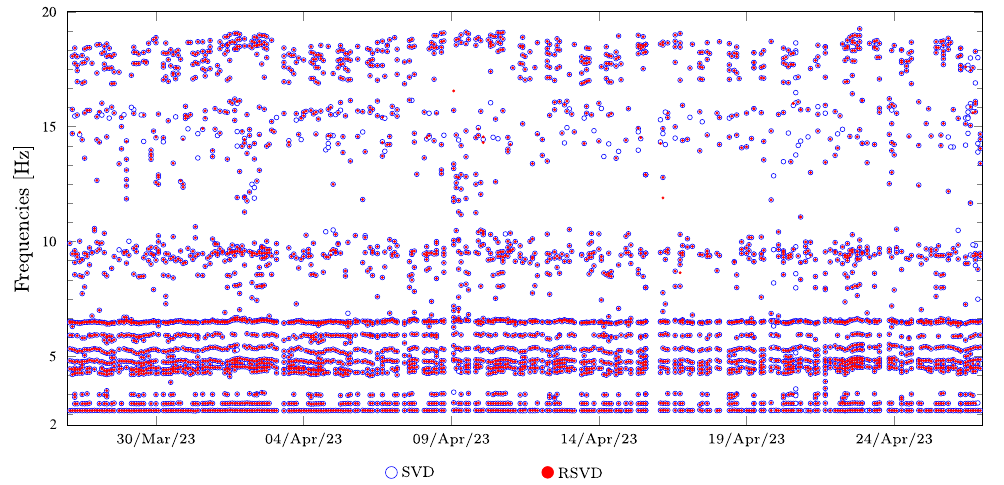}
   \caption{Identified modes between March 27$^{th}$ and April 27$^{th}$, 2023 of the San Faustino bridge.}    
    \label{fig:svd_rsvd_poles}
\end{figure}

The accelerations in the San Faustino bridge continuously recorded between March 27$^{\textrm{th}}$, 2023, and April 27$^{\textrm{th}}$, 2023, are used in this section to appraise the effectiveness of the proposed methodology for the long-term monitoring of this challenging bridge. As a first application, the long-term results obtained by using the classical SVD and the proposed RSVD-CoV-SSI are compared considering 2D stabilization diagrams. The modal identifications were performed considering model orders ranging from 250 to 400, analyzed in steps of 2. The time lag $\tau$ was chosen to be 3.775 s, corresponding to a time lag step $j_b$ of 76 as defined by Eq.~(\ref{eq:jb}), given the fundamental frequency $f_0$ of 2.64 Hz. The HC and SC applied in the computation of the stabilization diagrams included discarding poles with $\xi_i > 10\%$, MPC $\leq 60\%$, and MPD$\geq 50\%$. Additionally, relative differences in frequencies $f$, damping ratios $\xi$, and MAC values between different model orders greater than 2\%, 3\%, and 2\%, respectively, were considered. The size of the Toeplitz matrices $T$ was 8664, and the rank selected for the RSVD computation was 25\%, according to Eq.~(\ref{eq:empirical_formula}). The results of the continuous modal identification using  SVD and RSVD are reported in Fig.~\ref{fig:svd_rsvd_poles}. It can be noted that the results are perfectly overlapping, demonstrating the capability of the RSVD to substitute the SVD for modal identification purposes.

\begin{figure}[H]
\centering
   \includegraphics[scale=1.0]{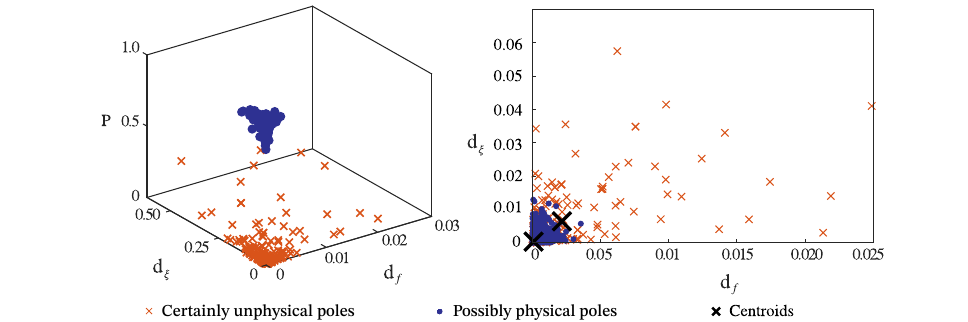}
   \caption{Results of the first stage of clustering of the San Faustino bridge.}    
    \label{fig:1stageclustering_San Faustino}
\end{figure}

\begin{figure}[H]
\centering
   \includegraphics[width=1\textwidth]{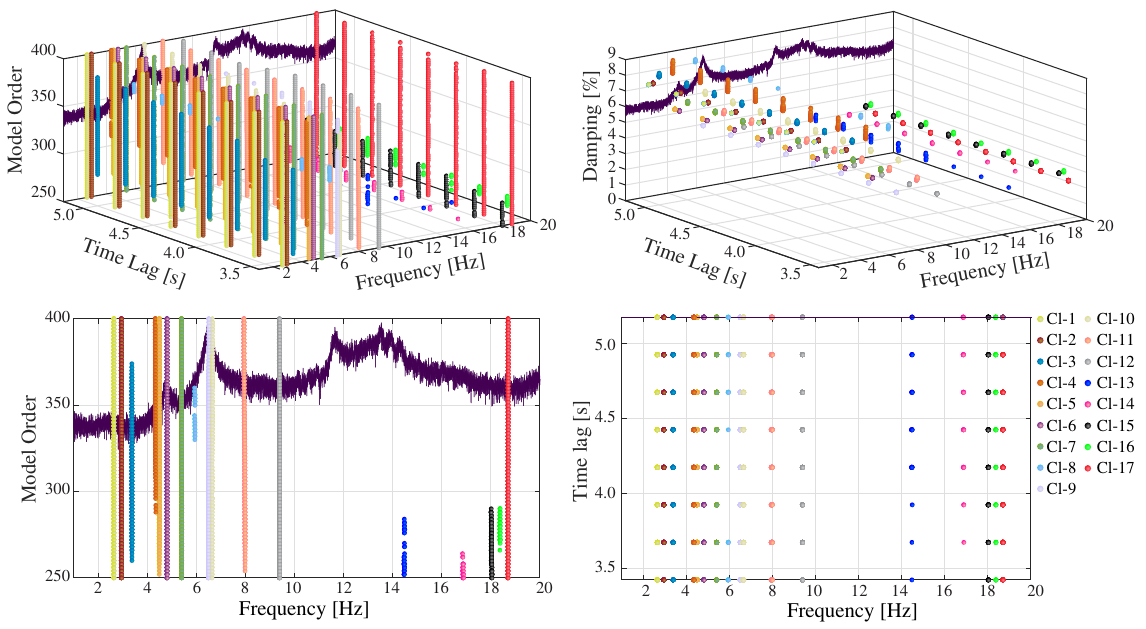}
   \caption{3D stabilization diagram of the San Faustino bridge (April 25\textsuperscript{th}, 2023, 2 p.m.). In the background (on the frequency-model order plane), the first singular value of the spectral density matrix is plotted for reference.}    
    \label{fig:3dstabdiag_sanfaustino}
\end{figure}

\begin{figure}[H]
\centering
   \includegraphics[width=1\textwidth]{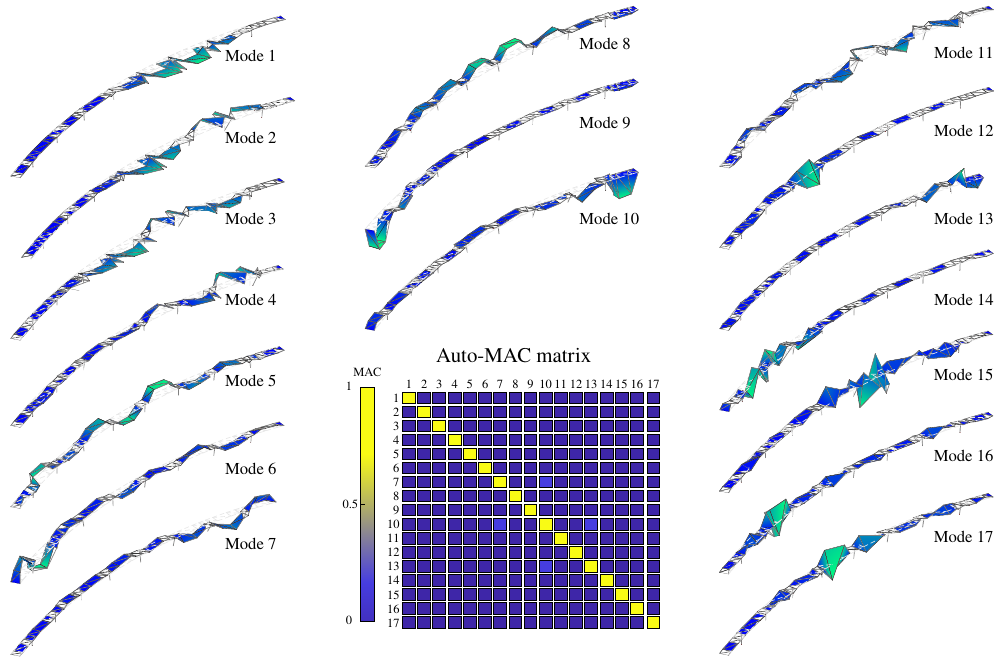}
   \caption{Mode shapes of the San Faustino bridge (April 25\textsuperscript{th}, 2023, 2 p.m.).}    
    \label{fig:modeshapes_SF}
\end{figure}

Afterward, the 3D stabilization diagrams introduced in Section \ref{section23} were used to continuously analyze the same datasets. Specifically, the same range of model orders was analyzed, and the considered time lags were between 3.43 s and 5.18 s with steps of 0.25 s, in accordance with Eq.~(\ref{eq:jb_3D}), that is $j_b$ values between 69 and 104. Therefore, the modal identification using the RSVD was performed for 8 different time lags around that specified by Reynders \textit{et al.}~\cite{reynders2008reference} in Eq.~(\ref{eq:jb}). Subsequently, the stability analyses of the poles were assessed on each pair of poles along the time lag and model order axes. According to the criteria defined in Eq.~(\ref{eq:stabcriteria3D}), the SC thresholds $\alpha_f$, $\alpha_\xi$, and $\alpha_{\textrm{MAC}}$ were slightly increased to 2.5\%, 5\%, and 3\%, respectively. This increase in the stability tolerances has been adopted due to the fact that altering the time lag can lead to more significant variations in the modal characteristics. Therefore, expanding the tolerances can be beneficial to prevent the exclusion of potentially physical poles. Regarding HC, the same criteria applied for the 2D cases were used. Then, the 2-stage clustering approach presented in Section~\ref{section23} was applied, considering the stable poles in the entire 3D space. The results of the first stage are reported in Fig.~\ref{fig:1stageclustering_San Faustino}. Finally, the hierarchical clustering algorithm was applied only to the stable poles retained after the first stage to find the physical mode shapes of the bridge. The 3D stabilization diagram of the San Faustino bridge is depicted in Fig.~\ref{fig:3dstabdiag_sanfaustino}, where 17 modes are reported, and the corresponding mode shapes are depicted in Fig.~\ref{fig:modeshapes_SF}. 
Slave equations were applied to plot the mode shapes, ensuring that nodes on opposite sides of the structure exhibit the same radial translation, since radial channels are present only on one side of the deck. In contrast, no slave equations were applied to the vertical nor the longitudinal channels. It should be noted that the reference parameters for the 2D identification were refined as described above. However, the complexity of the structure often prevents the generation of satisfactory 2D stabilization diagrams, as some modes are consistently missing compared to those that can be identified using 3D stabilization diagrams. 

\begin{figure}[H]
\centering
   \includegraphics[width=1\textwidth]{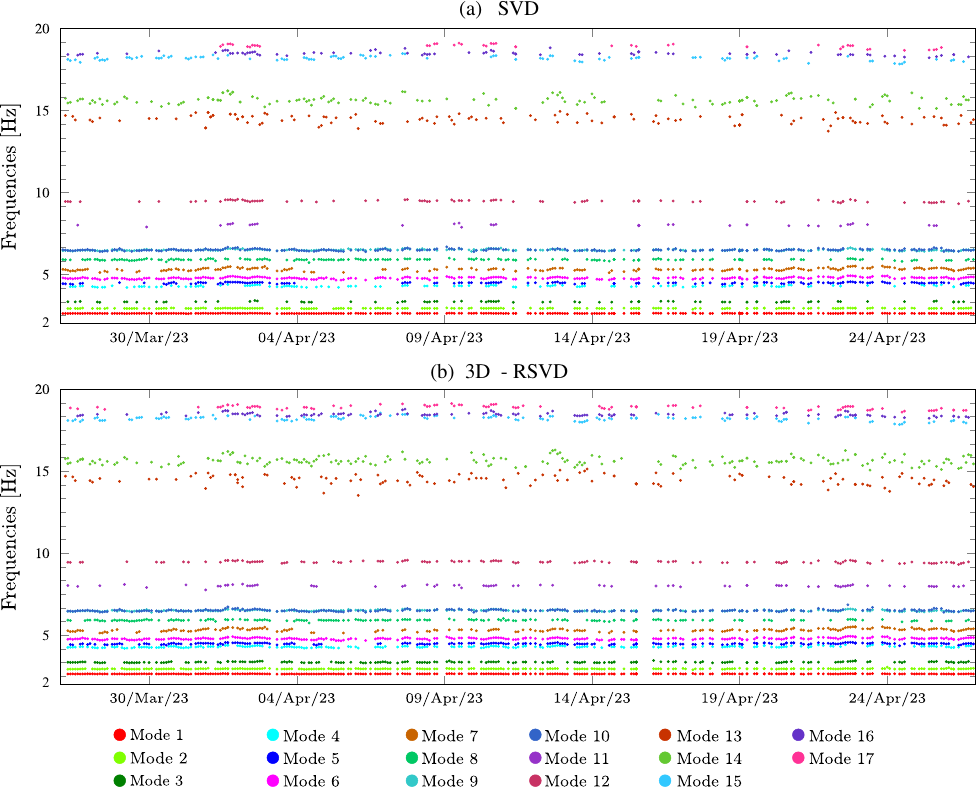}
   \caption{Tracking plots of the San Faustino Bridge using SVD for 2D Stabilization Diagram (a) and RSVD for 3D Stabilization Diagram (b).}    
    \label{fig:Tracking_SVD_3D}
\end{figure}

Finally, the modes derived from the 3D stabilization diagram were used to perform the tracking over the reference period. The main tracking parameters used in the analysis consider the maximum relative differences in terms of frequencies and MACs equal to 5\% and 15\%, respectively. The comparison between the modal tracking results using classical CoV-SSI against the proposed RSVD-CoV-SSI algorithm is depicted in Fig.~\ref{fig:Tracking_SVD_3D}. Significant improvements in tracking accuracy are observed when using 3D stabilization diagrams compared to the classical 2D approach. For the 2D stabilization diagrams, the RSVD results are close and comparable to those obtained with the SVD, while maintaining a significantly reduced computational burden. Specifically, using the computer defined in Section \ref{section31}, the average time required for a complete modal identification of the bridge using SVD is approximately 449 seconds, while for the RSVD with fixed time lag, it is about 204 seconds. In contrast, the 3D stabilization diagram with RSVD takes around 512 seconds. Therefore, although the 3D stabilization diagram takes roughly the same amount of time as the classical CoV-SSI, its efficiency in terms of modal identification accuracy is markedly improved. Additionally, Table~\ref{Tab4} reports the comparison between the identification success ratios for the 17 modes, defined as the ratio between the number of sets of time histories in which each mode was identified and the total number of time histories in the reference period, using both 2D stabilization diagrams with SVD and RSVD and 3D stabilization diagrams with RSVD.

\begin{table}[H]
\setlength{\tabcolsep}{3pt} 
\newcommand\Tstrut{\rule{0pt}{0,3cm}}         
\newcommand\Bstrut{\rule[-0.15cm]{0pt}{0pt}}   
 \footnotesize		
 \caption{ Comparison of the success ratios from modal tracking using classical CoV-SSI and the proposed RSVD-CoV-SSI. The columns labelled \textit{SVD} show results from classical CoV-SSI, while \textit{RSVD} and \textit{3D-RSVD} denote results from the CoV-SSI algorithm using RSVD, using classical stabilization diagrams and the proposed 3D stabilization diagrams, respectively. The reference modal properties are those identified by \textit{3D-RSVD} from the first acquisition record.}
 \vspace{0.1cm}
   \centering
   \begin{tabular}{cccccccp{0.2cm}ccccccc}
   \hline	
   \multirow{2}{*}{Mode} & \multicolumn{2}{c}{Ref. modal prop.} && \multicolumn{3}{c}{Success ratio [\%]} && \multirow{2}{*}{Mode} & \multicolumn{2}{c}{Ref. modal prop.}  && \multicolumn{3}{c}{Success ratio [\%]} \Tstrut\Bstrut\\
	 \cline{2-3}
   \cline{5-7}
	 \cline{10-11}
   \cline{13-15}
	&	$f_i$ [Hz]	&	$\xi_i$ [\%]	&&	\textit{SVD}	&	\textit{RSVD}	&	\textit{3D-RSVD}	&&		&	$f_i$ [Hz]	&	$\xi_i$ [\%] &&	\textit{SVD}	&	\textit{RSVD}	&	\textit{3D-RSVD}	\Tstrut\\
   \hline	
1	&	2.628	&	7.145	&&	99.0	&	98.7	&	100.0	&&	10	&	6.483	  &	5.950	&&	85.5	&	84.5	&	89.8	\Tstrut\\
2	&	2.972	&	7.182	&&	63.8	&	65.8	&	86.5	&&	11	&	7.994	  &	4.502	&&	11.5	&	10.9	&	23.4	\Tstrut\\
3	&	3.352	&	7.612	&&	26.0	&	25.0	&	49.0	&&	12	&	9.481	  &	3.352	&&	29.3	&	30.3	&	51.6	\Tstrut\\
4	&	4.339	&	7.868	&&	42.1	&	43.1	&	55.9	&&	13	&	14.677	&	3.085	&&	40.5	&	38.9	&	40.8	\Tstrut\\
5	&	4.475	&	5.268	&&	49.7	&	49.0	&	53.9	&&	14	&	16.546	&	2.927	&&	45.7	&	38.8	&	63.8	\Tstrut\\
6	&	4.827	&	5.111	&&	90.1	&	89.1	&	93.8	&&	15	&	18.100	&	3.251	&&	32.9	&	34.5	&	45.7	\Tstrut\\
7	&	5.353	&	5.552	&&	65.8	&	64.8	&	66.4	&&	16	&	18.291	&	3.314	&&	26.3	&	25.7	&	40.1	\Tstrut\\
8	&	5.903	&	8.184	&&	62.5	&	61.2	&	62.8	&&	17	&	18.850	&	2.690	&&	14.5	&	14.1	&	31.3	\Tstrut\\
9	&	6.474	&	3.908	&&	99.3	&	99.0	&	100.0	&&		  &		      &		&		&&		&		\Tstrut\\

   \hline
   \end{tabular}
   \label{Tab4}
\end{table}

\section{Conclusions}\label{section4}
This study investigates the use of randomized singular value decomposition (RSVD) for operational modal analysis (OMA), highlighting its potential to improve computational efficiency without sacrificing accuracy. RSVD approximates the SVD of a matrix using randomized sampling techniques, which reduces the computational complexity and hardware requirements compared to traditional SVD. Additionally, a novel automated OMA approach with reduced expert intervention is proposed exploiting the computational advantages of RSVD. Unlike classical CoV-SSI with fixed time lag for the computation of the correlation matrices, the time lag varied within a certain domain, assessing the stability of the resulting poles in 3D stabilization diagrams in the model order - frequency - time lag space. Afterwards, a two-step clustering approach combining fuzzy clustering and hierarchical clustering is adopted to automatically extract the physical poles of the structure. The effectiveness of RSVD is demonstrated with two case studies, including a theoretical 10-DOFs dynamic system and a challenging real-world application, the San Faustino Bridge in Italy, equipped with a dense instrumentation of 114 acceleration channels. The key findings of this work comprise:

\begin{itemize}
    \item RSVD demonstrated substantial savings in both computational time and memory usage, reducing computational time by more than 90\% and memory usage by more than 80\%.
    \item The fundamental form of the RSVD algorithm requires only one user-defined parameter: the rank $k$. An extensive parametric analysis has been carried out to establish and validate an empirical formulation of the minimum rank $\overline{k}$ required to perform the analysis while maintaining high accuracy in the identification.
    \item The computational savings from the RSVD implementation enabled the development of a 3D stabilization diagram for the modal identification of the San Faustino bridge, eliminating the need for fixing the time lag value for the computation of the correlation matrices. The reported results have demonstrated substantial improvements in the modal identification, allowing a better tracking of modal features with higher identification rates.
    \item A notable advancement is the potential for partial automation of the RSVD-based algorithm. The ability to automate the selection of key parameters, such as the time lag step and rank, using the empirical rules derived from this study can streamline the process of modal identification. This automation reduces the need for extensive manual tuning and expertise, making the algorithm more accessible and efficient for real-world applications.
\end{itemize}

In conclusion, the methodologies proposed in this work represent a significant advancement in the field of modal analysis. The adoption of RSVD not only makes it feasible to handle larger and more complex structures efficiently, but it also provides a more accurate and detailed understanding of modal characteristics by analyzing the stability with respect to time lag. The improved accuracy and reduced computational burden offer substantial practical benefits, particularly for monitoring networks of bridges and other large-scale infrastructure systems.

Future research could explore further optimizations in the RSVD algorithm to handle even larger datasets and more complex structures. Additionally, the integration of RSVD with other advanced techniques, such as machine learning and artificial intelligence for automated modal identification, could further enhance its applicability and performance. By addressing these aspects, the potential of RSVD in real-world applications can be expanded, leading to more efficient and effective modal analysis and structural health monitoring.

\section*{Acknowledgements}
\label{Sec:Acknowledgeoffunding}
The authors gratefully acknowledge funding from the Italian Ministry of University and Research through the PRIN Project “TIMING – Time evolution laws for IMproving the structural reliability evaluation of existING post-tensioned concrete deck bridges” (protocol no. P20223Y947). This study was also supported by FABRE  - “Research consortium for the evaluation and monitoring of bridges, viaducts and other structures” (www.consorziofabre.it/en) within the activities of the FABRE-ANAS 2021-2026 research program. The authors gratefully acknowledge ANAS S.p.a. for allowing access to the case study and to the monitoring data in the framework of the collaboration with FABRE. E. Garc\'{i}a Mac\'{i}as also acknowledges the support of the Spanish Ministry of Science and Innovation through the research projects ``BRIDGEXT - Life-extension of ageing bridges: Towards a long-term sustainable Structural Health Monitoring'' (Ref. PID2020-116644RB-I00) and ''SMART-BRIDGES - Monitorización Inteligente del Estado Estructural de Puentes Ferroviarios''. Any opinion expressed in the paper does not necessarily reflect the view of the funders.


\bibliographystyle{elsarticle-num}
\bibliography{Bibliography}

\end{document}